\documentclass[twocolumn,aps,superscriptaddress]{revtex4-1}
\pdfoutput=1

\usepackage[pdftex]{graphicx}
\usepackage{amsmath}
\usepackage{hyperref}
\usepackage{mathtools}
\DeclarePairedDelimiter\bra{\langle}{\rvert}
\DeclarePairedDelimiter\ket{\lvert}{\rangle}
\DeclarePairedDelimiterX\braket[2]{\langle}{\rangle}{#1 \delimsize\vert #2}

\begin{document}

\title{Unbiased (reference-free) phase field imaging for general
  optical fields \\ including phase discontinuities}

\author{Martin Berz} 
\affiliation{IFE Institut f\"{u}r Forschung und Entwicklung, 81675 Munich, Trogerstr. 38, Germany}

\author{Cordelia Berz} 
\affiliation{IFE Institut f\"{u}r Forschung und Entwicklung, 81675 Munich, Trogerstr. 38, Germany}

\date{\today}

\begin{abstract}
A new numerically and experimentally tested measurement method for the
local electrical light field including its phase is presented. The
method is based on Self Referencing Interferograms (SRI) such as
shearing interferograms. The complex electric field is the solution
vector of a linear equation with the pixel resolved interference term
$\overline E_2 E_1$ as a parameter. Linearization of the non linear
equations is achieved by using preknowledge in the intensity as
obtained by a conventional image detetctor. The resulting linear
equations are not based on any approximation, iterative perturbation
expansion etc.  but are exact.  The method is non iterative and stable
against noise for arbitrarily chosen test fields.  Allowed fields can exhibit
highly fluctuating amplitudes/phases on the pixel scale, areas of
vanishing amplitude and $\pi$ phase jumps.  The spatial resolution is
of pixel size. No reference beam and no diaphragms are used.  The new
method can be implemented as a fast, one shot per frame video
system. An outlook on the space resolved measurement of non classical
two photon states (including vacuum squeezed coherent states) is
given.
\end{abstract}

\maketitle


\section{Introduction}
\label{introduction}

Precise optical field measurements including the phase are a
prerequisite in many advanced optical systems such as adaptive optics,
metrology, 3D imaging, diffractive tomography or super resolution
microscopy~{\cite{yosh},\cite{born-wolf1991}}. The ideally needed
quantity is the complex amplitude of the optical field on a grid of
well defined space points. In the case of time dependent light fields, the
complex amplitude becomes a function of frequency. These data
completely characterize optical fields representing the maximum
available information for an imaging system. Quantum optical effects
can also be included (Section \ref{qd_optics}).

Technologies potentially suitable for such measurements are for
instance holography, Shack Hartmann detectors and SRI Self Referencing
Interferometers.

In classical holography, a reference beam is needed to generate the
hologram. A hologram contains the information about the complex field
amplitude. However, this holographic step is somewhat arbitrary since the
physical optical properties of the light field exist independently of
any reference beams. Hence, it should be preferred to extract the
holographic information or phase information without referring to a
reference beam. Thereby, all external perturbations in the reference
path e.g. vibrations, light path disturbances etc. are
eliminated. Such a measurement would measure the pure unbiased
information of the incoming field.

Avoiding the burdensome reference beam is particularly
necessary for advanced measurement situations where in principle no
'good' reference beam exists. This is likely the case for non linear
optical effects, for measurements over long macroscopic distances
('field measurements') or non-classical light (quantum optics).

In conclusion, it would be very advantageous to measure complex optical
amplitudes without depending on the arbitrariness of a reference beam.

Shack Hartmann detectors~\cite{yosh} are well established wavefront
detectors. The actually measured quantity is the locally averaged
angle of incidence of the incoming light. The average is taken over
the diameter of the individual micro lens. The angle of incidence is
synonymous to the averaged phase derivative over the micro lens
aperture. Each lens represents an individual measurement system. No
field comparison is made for light properties between different micro
lenses. As a result, no $\pi$ phase jumps can be detected. $\pi$ phase
jumps can occur in spatial light fields at points where the amplitude
of the field becomes zero, for instance in a standing wave. The
detector behind the micro lens might measure two overlapping spots or
no light at all. In any case, the micro lens does not give a valuable
signal which can be associated with the $\pi$ phase jump.

In SRI Self Referencing Interferometers the attempt is often
undertaken to locally generate  a flat zero mode optical field, which
serves as a reference beam~\cite{rhoadarmer}. This can be
achieved by a spatial band pass such as an illuminated pin hole or
single mode fiber leading to a loss of light though. Furthermore, the
reference generating process involves some intensity arbitrariness of
the reference field. This introduces noise sources in the overall
process. Thus, this kind of SRI is not a really satisfying solution
yet.

Another approach is to measure shearing interferograms and disentangle
the electrical fields by solving the associated non linear
equation~\cite{falldorf}. The method is mainly developed for shearing
interferometers and uses an iterative approach which evidently must be
damped by a smoothing term~\cite{falldorf}. In general, convergence
and stability is a big challenge in phase reconstruction.

A new method is introduced here which overcomes many limitations of
the methods mentioned above. The associated new device is technically
called HOLOCAM, an artificial word created from HOLOgraphy and
CAMera. The device is holographic since it uses a holographic type
evaluation of holograms or interferograms and it measures something
'holographic', the complex amplitude. The device is also a camera such
as a CCD camera which registers the locally resolved light intensity
on a flat detector. The HOLOCAM does something similar but the
registered quantity is the complex amplitude field.  Both the CCD and
the HOLOCAM do not need a reference beam. The phase is obtained by
solving a linear equation. The solution space of linear equations
can be completely characterized. Hence, there is no convergence
problem. Besides, it will be shown in Section \ref{sensitivity} to
\ref{num_res} that the solution is stable against noise. This is a
remarkable result for phase retrieval which otherwise is often
unstable.

As will be shown in Section \ref{ref_BEX} the HOLOCAM can directly
resolve arbitrary fields including fields with $\pi$ phase jumps.

The HOLOCAM method has been experimentally tested using a lateral
shearing interferometer.

\section{Derivation of the fundamental linear field equation}
\label{sec_fund_equ}

\begin{figure}
\begin{center}
\includegraphics[scale=0.45]{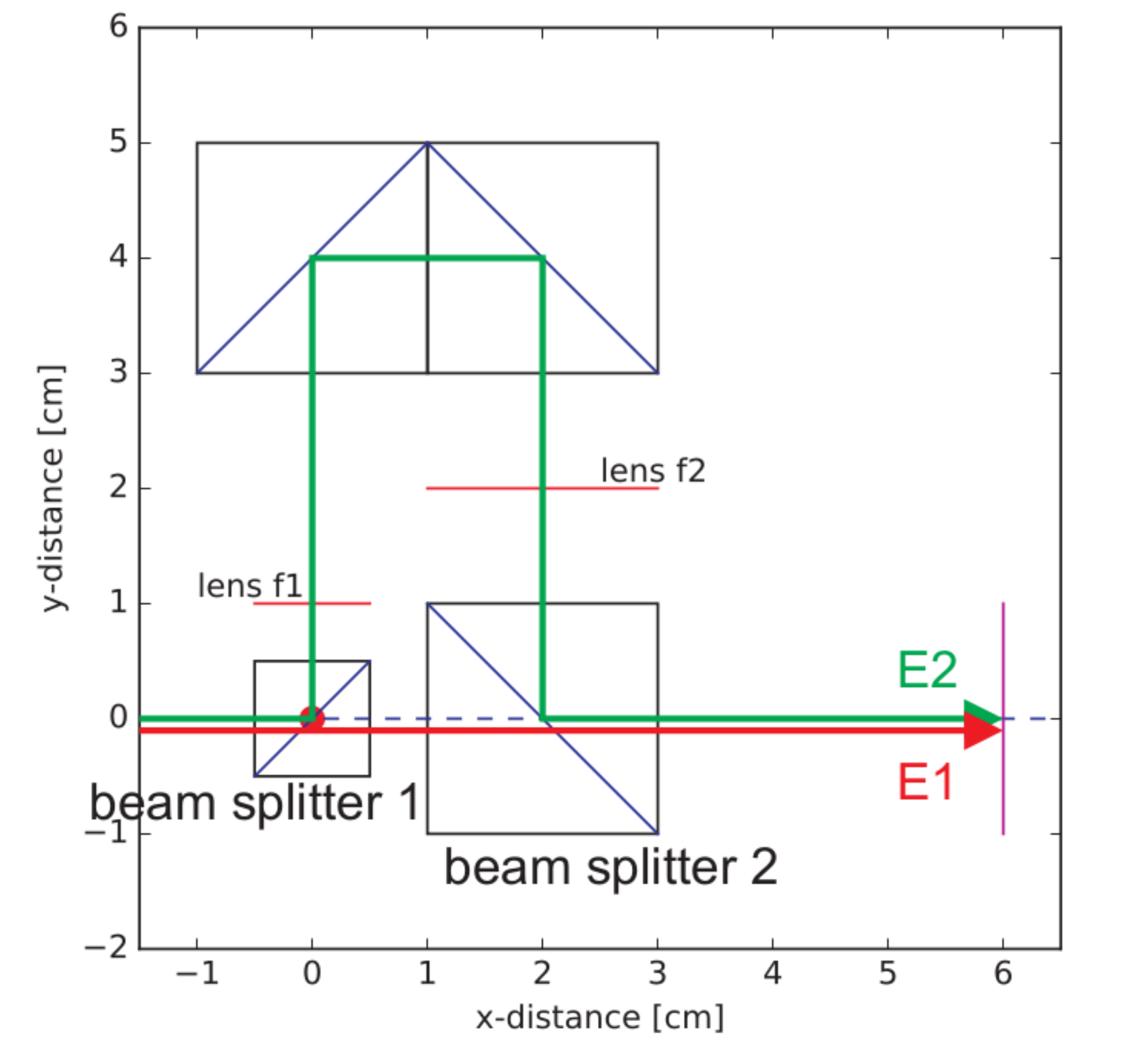}
\caption{\label{bex_setup} The HOLOCAM in a beam expander setup. The
  light enters by the left side and is splitted at beam splitter 1 in
  two beams. One beam ($E1$) passes straightly through the
  interferometer and hits an imaging plane at the right. The second
  beam passes by branch 2. Branch 2 contains 2 lenses which can be
  arranged such as to form a beam expander.}
\end{center}
\end{figure}

Figure \ref{bex_setup} shows an exemplary HOLOCAM device. It consists
of a Mach Zehnder type interferometer with two unequal branches. The
upper branch (branch 2) is longer and contains, in this example, two
lenses. The light enters from the left, is then splitted into two
beams and these two beams subsequently come to interference on a
screen on the right. The two lightfields are called $E_1$, $E_2$ being
propagated by branch 1,2 respectively.

The intensity $IN$ that is measured on the screen can be expressed as

\begin{equation}
\label{eq:basic_IF}
IN(\vec x_s) = \overline {(\vec E_1(\vec x_s) + \vec E_2(\vec x_s))} \cdot
(\vec E_1(\vec x_s) + \vec E_2(\vec x_s))
\end{equation}

Here and in the following $ \cdot $ denotes the scalar product of the
field vectors on every pixel. In Equation (\ref{eq:basic_IF}) it is
the scalar product between the 3D E-vectors.

$E_1(\vec x_s)$,$E_2(\vec x_s)$ are the vectorial first and second
electric field, respectively. $\vec x_s$ is the location where the
field is measured, s being some index of location, such as an index of
a pixel. In Figure \ref{bex_setup} the pixel is part of the
plane where the fields $E1$ and $E2$ overlap on the right. To simplify
the notation, the time variable $t$ (respectively $\omega$) is
supressed. The particular meaning should be clear by context.  In the
following, $\vec E_1(\vec x_s)$ stands for a quasi monochromatic field
with angular frequence $\omega$, yielding $\vec E_1(t, \vec x_s ) =
\vec E_1(\omega, \vec x_s ) e^{i \omega t} $.  Equation \ref{eq:basic_IF}
is actually independent of the factor $e^{i \omega t} $. Dropping the
vector notation and using the abbreviation $E_1(\vec x_s) =: E1_{s}$
Equation \ref{eq:basic_IF} becomes

\begin{equation}
\label{eq:basic_IF2}
IN_{s} = \overline {( E1_s + E2_s)} *  {( E1_s + E2_s)} 
\end{equation}

or

\begin{equation}
\label{eq:basic_IF3}
IN_{s} = \mid E1_s \mid^2 + \mid E2_s \mid^2  + 2 \ Re ( \ \overline{E2_s} * E1_s \ )
\end{equation}

or further

\begin{equation}
\label{eq:basic_IF4}
IN = \mid E1 \mid^2 + \mid E2 \mid^2  + 2 \ Re ( \ \overline{E2} * E1 \ )
\end{equation}

Here and in the following $*$ denotes a point wise multiplication,
for instance two pixel vectors $\{x_s\}$,$\{y_s\}$ are $*$ multiplied to yield
a new vector $\{z_s\}$ of the same dimension with $x_s * y_s = z_s$
for all pixel indices $s$. '*' is essentially different from a scalar
product such as '$\cdot$' since it does not include summation. In most
cases, the indices $s$ enumerate the pixels of the
detector. In some cases (such as in Equation (\ref{eq:basic_IF2})) the
running pixel index 's' is still kept but '*' is used. This is actually a
redundant expression for the fact that the multiplication is
individually performed on every pixel without summation.

Equation (\ref{eq:basic_IF3}) is the well known interference
expression. $\mid E1_s \mid$, $\mid E2_s \mid$ are the amplitudes of
the electric field from branch 1,2 respectively. These quantities can
be measured by conventional methods. Subtracting this background in
Equation (\ref{eq:basic_IF3}) shows that
$Re ( \ \overline{E2_s}*E1_s\ )$ can be measured. 

Using known technologies such as 'phase shifting' or 'carrier phase'
not only $Re ( \ \overline{E2_s}*E1_s\ )$ but the full complex
interference term can be measured. This quantity is called IF.

\begin{equation}
\label{eq:basic_IF5}
IF_{s} = \overline{E2_s} * E1_s 
\end{equation}

$IF_{s}$ is a complex field given for all index positions
$s$. $IF_{s}$ will be called a hologram in contrast to $IN_s$ which is
called the inferogram.

In the following, it will be assumed that an appropriately designed
interferometer is used. The interferometer should have the property
that the set of $ \big\{ E2_s \big\} $ for all indices $s$ is determined
by $ \big\{ E1_t \big\} $ for all indices t. A simple example for such
an interferomter is given in Section \ref{ref_BEX}. Furthermore, the index
for polarization will be suppressed. Therefore, polarization effects will
be neglected. In many cases this is justified. If needed all equations
can be generalized straightforwardly to include explicit
polarization. The assumption about the polarization is particularly
correct if $E_1$ and $E_2$ have the same polarization. Then $E_1$
becomes a scalar quantity and the '$\cdot$' operation in Equation
(\ref{eq:basic_IF}) can be suppressed.

Subsequently, a mapping U from fields in branch 1 to fields in branch
2 will be defined.The mapping is defined on the pixels {$x_s$} of the
detector.

\begin{equation}
\label{eq:basic_U}
E2_{s} = \sum_{t} U_{s,t} E1_t =: U(E1) =: U E1 
\end{equation}

For matrices A,B the term A B (without $\cdot$ or $*$ ) stands, as
usual, for the matrix multiplication of two matrices. B might be a
vector (such as E1) in which case A B stands for the matrix
multiplication between A B yielding a new vector. In this case, A
stands for a linear mapping (such as U).

The linear mapping U: $E1 \rightarrow E2$ in Equation
(\ref{eq:basic_U}) is actually a field propagation. It can be
calculated by solving Maxwell's equations. In the simplest case
of a Mach-Zehnder interferometer it represents a back propagation of
field E1 from the detection plane given by the locations $\big\{ x_t
\big\}$ to a plane before the first beam splitter and thereafter a
forward propagation from this plane to the detection plane given again
by the locations $\big\{ x_s \big\}$. The back propagation is along
the first branch of the interferometer and the forward propagation is
along the second branch of the interferometer. Hence, U is in fact
composed of two propagations.

The propagation could obviously show diffraction effects. Moreover, it
is possible to choose appropriate focal lengths $f_1$, $f_2$ in Figure
(\ref{bex_setup}) so that U becomes a point mapping.

The mapping U can only be well defined if the Nyquist Shannon 
criterium is respected. This means that the $\pi$ phase variations of
E1, E2 must not be narrower than the $x_s$ spacing on the
detector.

Care must also be attributed to signal run off through the borders of
the representation space $\big\{ x_s \big\}$. The run off of signal
strength through the borders might well happen in realistic
situations. This signal strength does not affect the result though as
it is not reflected. The run off of signal strength can be treated
physically and numerically exactly. In many cases, it is not too difficult to
choose an appropriate design for which U is a well defined physical
map. U is usually injective but not necessarily unitary. Examples for
U will be given in Section \ref{ref_BEX}.

For the following section, it is important to state that U represents
the physical relationship between E1 and E2 and that U is a property
of the measuring device. It does not depend on the incoming
field. Having sufficient knowledge of the used interferometer U can be
calculated even though this might involve some calibration steps. Once
these calibration steps have been properly performed the matrix U is
determined and fixed for subsequent measurements of unknown fields.

For completeness, it is mentioned that the introduced device is
conceived for a certain group of incoming fields. The allowed group of
incoming fields is given by an accepted optical field of view and an
allowed incoming field diameter. These are characteristics well
known from other optical measuring devices. Therefore, the restrictions
in the applicability of U are not particular but usual for optical
devices.

Under these circumstances the following equation holds

\begin{equation}
\label{eq:if_1}
IF_s = \overline{E2_{s}} * E1_{s} = E1_{s} * \sum_{t} \overline {U_{s,t}
  E1_t}
\end{equation}

Whenever needed $E2$ can be expressed by $E1$. As a consequence,
Equations (\ref{eq:basic_IF4}) and (\ref{eq:if_1}) are nonlinear
equations for the complex field $E1$.  This is normally expressed as an
E1 optimization problem~\cite{falldorf}:

\begin{equation}
\label{eq:opti}
\left\| IF_s - E1_{s} * \sum_{t} \overline {U_{s,t} E1_t} \right\| \rightarrow \min
\end{equation}

Equation (\ref{eq:opti}) is non linear and probably unstable in the general
case.

A new approach is proposed in the following which avoids the
convergence problem. For this Equation (\ref{eq:if_1}) is multiplied
by $\overline{E1_{s}}$ yielding

\begin{equation}
\label{eq:if_2}
IF_s * \overline{E1_{s}} = \mid E1_{s} \mid^2 \ * \sum_{t} \overline {U_{s,t}} \ \overline { E1_t} 
\end{equation}

\begin{equation}
\label{eq:fund_eq}
IF * \overline{E1} =  \mid E1 \mid^2  \ * \ \overline {U E1} 
\end{equation}

In Equation (\ref{eq:if_2}) the term $\mid E1 \mid^2$ can be
determined by a conventional intensity measurement. In the following,
it is assumed that $\mid E1 \mid^2$ is already known. Having this
pre-knowledge in $\mid E1 \mid^2$ Equation \ref{eq:if_2} becomes a
linear equation in the unknown complex field $\overline { E1}$. 

Taking the conjugate complex of Equation (\ref{eq:if_2}) this can be
reformulated as an equation for $E1$. Equation (\ref{eq:fund_eq}) is
a compact formulation in matrix notation called the 'fundamental
equation' in this context. To the best available knowledge this
reformulation of the phase problem has remained unnoticed up to now.

The fundamental Equation (\ref{eq:if_2} , \ref{eq:fund_eq} )
represents an exact, simple and linear equation for the complex
amplitude of field $E1_s$. The measured quantities $\mid E1_s \mid^2$
and $IF_s$ enter the equation as coefficients. The solution space of
linear equations is well known. Hence, the difficulties arising from Equation
(\ref{eq:opti}) are avoided. In the subsequent sections (Sections
\ref{sensitivity}, \ref{ref_BEX}, \ref{preconditioning}), the
uniqueness and stability of Equation (\ref{eq:fund_eq}) will be
analyzed for properly designed devices.

\section{Sensitivity of the fundamental equation}
\label{sensitivity}

For stability analysis a calculation of the functional derivative
would be adequate. This is a high dimensional object equivalent to a
perturbation expansion as known from quantum
mechanics~\cite{messiah}. The notation of Messiah will be used in the
following. The approach allows us to understand the mechanisms
responsible for the perturbation (error) propagation.

A perturbation term is introduced in Equation (\ref{eq:if_2}) and the
new solution is expanded as a function of the perturbation term.  The
linear operator $H_0$ of the unperturbed problem is introduced:

\begin{equation}
\label{eq:pert1}
0 = H_0 ( \overline{E1}) := \mid E1 \mid^2  \ * \ \overline {U} ( \overline{E1}) - IF *  \overline{E1}  
\end{equation}

The exact solution of the unperturbed problem corresponds to a zero
eigenvalue.

Secondly, a perturbed operator H is defined

\begin{equation}
\label{eq:pert2}
H  \overline{E^p}  := H_o \overline{E^p} + \delta IF*  \overline{E^p}  
\end{equation}


'p' stands for perturbed. ${E^p}$ can no longer be expected to be an
eigenvector to the eigenvalue zero. In fact, some decision has to be
made concerning how to define a solution for the perturbed
problem. The option chosen here is to look for the adiabatic
eigenvector as a function of the perturbation. This is equivalent to
looking for the lowest eigenvalue of the perturbed problem Equation
(\ref{eq:pert2}). Another possible approach would be to find a
least square solution of (\ref{eq:pert2}). The latter is equivalent to the
lowest singular value solution of (\ref{eq:pert2}).

$\delta IF$ represents measurement errors in the determination of the
hologram $IF$. As $IF$ also $\delta IF$ is an array of complex
numbers.

$E1$ denotes an eigenvector of the unperturbed system for the
eigenvalue zero. Contrary to quantum mechanics, neither $H_0$ nor
$\delta IF$ are self adjunct operators. However, without too much loss
of generality we assume a non degenerate spectrum of $H_0$ and
consequently a complete set $\bra{n}$, $\ket{n}$ of left and right
hand eigenvectors, defined as

\begin{equation}
\label{eq:hdef1}
H_o \ket{n} = C_n \ket{n}
\end{equation}

and 

\begin{equation}
\label{eq:hdef2}
\bra{n} H_o  = C_n \bra{n}
\end{equation}

respectively. Normalization is chosen such as 

\begin{equation}
\label{eq:hdef3}
\braket{m}{n}   = \delta_{m,n}
\end{equation}

and 

\begin{equation}
\label{eq:hdef4}
\lVert \ket{n} \rVert = 1
\end{equation}

holds.

$C_n$ are the complex eigenvalues of $H_0$.

All vectors are tupels in $s$ - index space of pixels. The bra -
vectors $\bra{n}$ are not normalized with respect to the complex
scalar product $\sum_s \overline{a_s} b_s $ but with respect to the
Kronecker product Equation (\ref{eq:hdef3}).

Using this notation the perturbation theory known from quantum
mechanics can be used~\cite{messiah}. Therefore, an expansion parameter
$\lambda$ is introduced which will be set to $\lambda = 1 $ later.

\begin{equation}
\label{eq:pert5}
\overline{E1^p} (C_0 + \delta C_o) = H  \overline{E1^p}  = H_o \ \overline{E1^p} + \lambda \ \delta IF *  \overline{E1^p}  
\end{equation}

and the eigenvalues and eigenvectors become

\begin{equation}
\label{eq:pert6}
C_0 + \delta C_o = \zeta_0 + \lambda \zeta_1 + \lambda^2 \zeta_2 + ... + \lambda^n \zeta_n + ...
\end{equation}

$C_0 + \delta C_0$ is the eigenvalue which evolves from the lowest
eigenvalue $C_0$ of the unperturbed system. If no other perturbations
are present $C_0 $ and $\zeta_0$ equal zero (for the fundamental
equation).

\begin{equation}
\label{eq:pert7}
\overline{E1} + \delta \overline{E1} = \ket{e_0} + \lambda \ket{e_1} + \lambda^2 \ket{e_2} + ... + \lambda^n \ket{e_n} + ...
\end{equation}

By definition the relation $\overline{E1} = \ket{e_0}$
holds. $\overline{E1^p}$ is the eigenvector to
the eigenvalue $C_0 + \delta C_0 = \delta C_0 $ of the perturbed
problem.

According to Messiah\cite{messiah} the operator $\frac{Q}{a}$ is introduced

\begin{equation}
\label{eq:hdef8}
\frac{Q}{a} = \sum_{n>0} \frac{\ket{n}\bra{n}}{C_0 - C_n} 
\end{equation}

This yields the following iterative solution

\begin{equation}
\label{eq:hdef9}
\zeta_n = \bra{\overline{E1}} \delta IF  \ket{e_{n-1}} 
\end{equation}

\begin{equation}
\label{eq:hdef10}
\ket{e_n} = \frac{Q}{a} \Big[ ( \delta IF - \zeta_1) \ket{e_{n-1}} - \zeta_2 \ket{e_{n-2}} - ... - \zeta_{n-1} \ket{e_1}  \Big]
\end{equation}

In the first order the following formula holds

\begin{equation}
\label{eq:1st_order}
\delta \overline{E1} \doteq \ket{e_1} = \frac{Q}{a} \Big[ ( \delta IF - \zeta_1) * \overline{E1} \Big]
\end{equation}

Since  $\frac{Q}{a} \overline{E1} = 0$ by definition this can be further simplified to

\begin{equation}
\label{eq:1st_order_compact}
\delta \overline{E1} \doteq \ket{e_1} = \frac{Q}{a} \ (\delta IF * \overline{E1})
\end{equation}

Equation (\ref{eq:1st_order_compact}) is the desired simple small
scale formula describing error propagation in the fundamental Equation
(\ref{eq:fund_eq}).

In the next two sections (Sections \ref{ref_BEX} to \ref{num_res})
this formalism will be used to study a simple reference case. The error
propagation matrix in Formula (\ref{eq:1st_order_compact}) is still a high
dimensional object. Therefore, it is favorable to condense the
result to a single key number further. $\delta \overline{E1}$ expresses how
much the solution of the perturbed problem deviates from the exact (or
'correct') solution. Hence, it is appropriate to calculate the standard
deviation for $\delta \overline{E1}$.  This is the vector norm of
$\delta \overline{E1}$ divided by $\sqrt{N_{det}}$, $N_{det}$ being
the number of detection points.

The standard deviation is

\begin{equation}
\label{eq:sigma}
\sigma := \sqrt {\frac {\sum_s \lvert \delta E1_s \rvert^2 }{N_{det}}} = \frac{\lVert \delta \overline{E1} \rVert}{\sqrt{N_{det}}}
\end{equation}

The result still depends on the scaling of E1. The reference for
the signal strength is the average of E1. Thus, as usual in
signal-to-noise figures the expression for the standard deviation (the
'noise') is divided by the mean signal strength $av(\lvert E1 \rvert)$.

\begin{equation}
\label{eq:av_E1}
av(\lvert E1 \rvert) := \frac {\sum_s \lvert E1_s \rvert }{N_{det}} 
\end{equation}

This yields the following quality (or S/N) parameter called $\sigma_{\delta E1}$:

\begin{equation}
\label{eq:sigma_E1} 
\sigma_{\delta E1} = \frac{\lVert \delta \overline{E1} \rVert}{\sqrt{N_{det}}}\frac{1}{av(\lvert E1 \rvert)}
\end{equation}

For $\delta IF$ a similar measure can be found:

\begin{equation}
\label{eq:sigma_IF} 
\sigma_{\delta IF} = \frac{\lVert \delta IF \rVert}{\sqrt{N_{det}}}\frac{1}{av(\lvert IF  \rvert)}
\end{equation}

The following characteristic number for the error propagation in the system can be defined:

\begin{equation}
\label{eq:sigma_ratio} 
A_{error} := \frac{\sigma_{\delta E1}} {\sigma_{\delta IF}}  = \frac{\lVert \delta \overline{E1} \rVert}{\lVert \delta \overline{IF} \rVert} \frac{av(\lvert IF \rvert)}{av(\lvert E1 \rvert)}
\end{equation}

This ratio is called the error amplification factor.

For the phase error a normalized standard deviation can also be defined: 

\begin{equation}
\label{eq:sigma_phi} 
\sigma_{\delta \phi_1} = \frac{\lVert \delta \phi_1 \rVert}{\sqrt{N_{det}}}
\end{equation}

No reference to the signal size is necessary since the variation of the
phase error is always limited by $2\pi$.

There are two possible origins of $ \delta IF$, either a, for
instance, gaussian random noise source or a projection of the signal
IF on a discrete integer space given by the Bit space of the
detector. The difference to the unprojected value is the error caused
by the detector quantization. For the calculation of the error
amplification factor the origin of $ \delta IF$ is not of primary
importance since every noise source is condensed to a $\sigma_{\delta IF}$.
Nevertheless, the process of error generation slightly influences the
value $A_{error}$. In the following, the type of noise generation will
be noted for numerical results yielding two cases, either gaussian
noise or quantization noise.

The following quantity $A\phi_{BIT-error}$ for quantization noise is
particularly interesting because it is independent of any normalization:

\begin{equation}
\label{eq:phi_error} 
A\phi_{BIT-error} := \frac{\sigma_{\delta \phi}} {quantization} :=
\frac{\lVert \delta \phi_1 \rVert}{ \sqrt{N_{det}} \ 2^{-n_{bit}} }
\end{equation}

$n_{bit}$ is the Bit number of the signal IF, i.e. $n_{bit} = 10$
means that the signal IF is  projected on a 10 Bit number. The difference between the 'exact' value IF and the projected value is just $\delta IF$. 

In the subsequent Section \ref{ref_BEX}, a simple HOLOCAM system will
be designed ('the beam expander HOLOCAM'). In the Section
\ref{num_res} the error amplification factor $A_{error} $ and the
lowest eigenvalue separation $\Delta_{min} C$ will be calculated for
the 'the beam expander' HOLOCAM.



\section{Design of a simple HOLOCAM system, the beam expander HOLOCAM}
\label{ref_BEX}

One of the simplest systems to study is a one dimensional system with
a Mach Zehnder interferometer (Figure \ref{bex_setup}). The system
could for instance be built using a linear detector array. Hence, the
terminus 'one-dimensional' only concerns the detector. In the second
branch, a lens system is installed in such a way that the detector
plane self mapped by U becomes a conjugate plane. This means a field
point in the detector plane back propagated by branch one and then
forward propagated by branch two (including the mapping by the lens
system) becomes again a point image in the detector plane. The mapping
U is thus a point mapping. An ideal system is assumed neglecting any
aberrations in the optical system. If this was not the case the
aberrations could be included in U. This is without relevance for the
stability analysis though.

The expansion ratio of the beam expander is called $\gamma$ and if not
otherwise stated $\gamma = 2 $ is used. $\gamma = 2 $  means
that the image of the detection plane is expanded by a factor of two.

The point mapping U applied in the discrete pixel space implies some
interpolation since the mapping $0, 1 ,2, 3 ..$ to $0,2,4,6,.. $ needs
to be accomplished by field values of $E2$ at intermediate pixels at
$1,3,5,..$. This does not imply any loss of information. Interpolation
will be further discussed in Section \ref{discretization}.

A simplification is made by assuming a one dimensional detector
array. This systems can actually be built. The detector space not the
propagation space is one dimensional. This assumption does not affect
the results. Besides, general 2D problems will be studied later on in
Section \ref{num_res}.

In the case of 50 detector points and for a selected pixel range from 20
to 30 this could be representetd by the following U matrix:

\begin{equation}
\left[\begin{matrix}0.0 & 0.0 & 0.5 & 0.5 & 0.0 & 0.0 & 0.0 & 0.0 & 0.0 & 0.0\\0.0 & 0.0 & 0.0 & 1.0 & 0.0 & 0.0 & 0.0 & 0.0 & 0.0 & 0.0\\0.0 & 0.0 & 0.0 & 0.5 & 0.5 & 0.0 & 0.0 & 0.0 & 0.0 & 0.0\\0.0 & 0.0 & 0.0 & 0.0 & 1.0 & 0.0 & 0.0 & 0.0 & 0.0 & 0.0\\0.0 & 0.0 & 0.0 & 0.0 & 0.5 & 0.5 & 0.0 & 0.0 & 0.0 & 0.0\\0.0 & 0.0 & 0.0 & 0.0 & 0.0 & 1.0 & 0.0 & 0.0 & 0.0 & 0.0\\0.0 & 0.0 & 0.0 & 0.0 & 0.0 & 0.5 & 0.5 & 0.0 & 0.0 & 0.0\\0.0 & 0.0 & 0.0 & 0.0 & 0.0 & 0.0 & 1.0 & 0.0 & 0.0 & 0.0\\0.0 & 0.0 & 0.0 & 0.0 & 0.0 & 0.0 & 0.5 & 0.5 & 0.0 & 0.0\\0.0 & 0.0 & 0.0 & 0.0 & 0.0 & 0.0 & 0.0 & 1.0 & 0.0 & 0.0\end{matrix}\right]  
\end{equation}

This is called 'geometric interpolation', $U_x$.

The same kind of interpolation can also be made in Fourier
space. Assuming the signal has an upper cup in the size of the Fourier
coefficients yields another unambiguous interpolations method. Here,
the Fourier representation of the field, which is a continuous
function, is evaluated at intermediate points. This interpolation will
be called 'Fourier interpolation' yielding $U_f$.

From the preceding, it is obvious that some points are, as a
consequence of discretization, calculated by interpolation. Even in
this simple case U cannot be represented by a strict point
mapping. The latter would be necessary if one wished to take a
pointwise logartithm of Equation (\ref{eq:basic_IF5}) as done in
reference \cite{malacara}. The error becomes small for very smooth
fields with little or no amplitude variation which is not the case for
general fields though. Logartithm of Equation (\ref{eq:basic_IF5}) is
consequently no option for solving the non linear Equation
(\ref{eq:opti}).

In the following, the solution of the exact Equation
(\ref{eq:fund_eq}) is further analyzed. This is done by introducing
random fields for which IF is calculated and for which the original
field must then be recovered solving the fundamental Equation
(\ref{eq:fund_eq}).

\begin{figure}
\begin{center}
\includegraphics[scale=0.5]{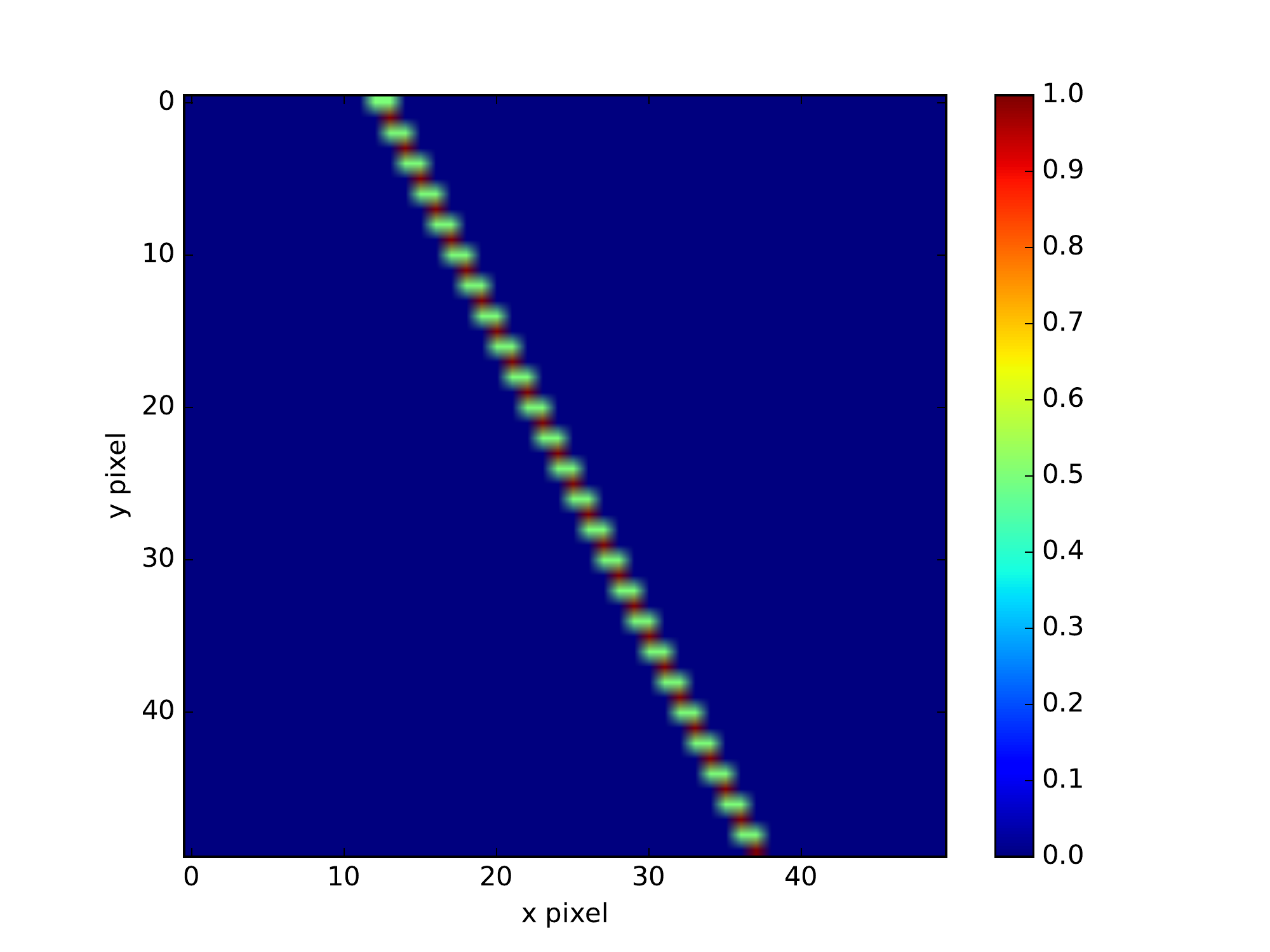}
\caption{\label{fig1}
The matrix Ux for the beam expander setup.
}
\end{center}
\end{figure}
 
Hence, for further analysis a random field $E_1$ must be defined.

\begin{equation}
\label{eq:e1_random}
E1(x_s) = a (\varsigma_s +  i \ \varrho_s)
\end{equation}

$\varsigma_s$,$\varrho_s$ being continuous uniform random numbers
equally distributed in the half open interval $[-1.0, +1.0)$. 'a' is a
  real normalization constant. Adjacent points are absolutely
  uncorrelated and the phase relation between adjacent points is also
  random in the range $[-\pi, \pi)$.  The unwrapped phase corresponds
    to a one dimensional random walk with a maximal jump length of
    $\pi$. This is the maximum variation according to the Nyquist
    Shannon theorem.

The random field in Equation (\ref{eq:e1_random}) has by design a flat
Fourier spectrum up to the maximum k-vector ($k_{max} = \frac{2 \pi}{2
  \Delta_{pix}}$. $IF$ is a quantity which is in real space a point
wise product of $E1$ and $E2$. In order to investigate this influence
further $E1$ is not only used in its raw version of Equation
(\ref{eq:e1_random}) but also Fourier filtered by a high frequency cut
off in the Fourier spectrum of $E1$. If the cut off is chosen to be
for instance $0.5 k_{max}$ the input field $E1$ is said to be averaged
over two pixel. This is labeled by $n_{av}=2$. $n_{av}=1$ means 'no
averaging'. Other magnitudes of averaging are treated analogously.

According to Section \ref{sensitivity} the noise sensitivity of the
correct solution is determined by the error amplification factor
$A_{error}$, the Bit error amplificationfactor $A\phi_{BIT-error}$ and
other quantities definined in Section \ref{sensitivity}.
    
In the case of the determination of the Bit error
($A\phi_{BIT-error}$) no further random quantity is needed. Otherwise,
$\delta IF$ must be defined explicitly by another random field. For
this, a gaussian random field with a standard deviation
$\sigma_{\delta IF}$ is chosen.

In brief, the terrain is prepared for the numerical runs. As assumed
from the beginning the amplitude $\mid E1 \mid$ is known from a
conventional intensity measurement.

Before doing the actual runs though preconditioning of the fundamental Equation
(\ref{eq:fund_eq}) will be discussed in the following section.

\section{Preconditioning of the fundamental equation}
\label{preconditioning}

E1 is an eigenvector of the 'exact' fundamental equation, expressed as
$H_0$. Due to measurement errors (perturbations) not $H_0$ but some
$H$ is measured or known. The question is what kind of property of $H$
has to be determined to recover the best approximation of E1 ?

In Section \ref{sensitivity} this has been done by looking for
the eigenvector of the lowest eigenvalue. Although this is a valuable
strategy it is not necessarily the best one.

Apart from that, Equation (\ref{eq:fund_eq}) can also be solved by
minimizing the residual ('least square solution'). This is equivalent
to the solution vector for the lowest singular value ('SVD -
decomposition').

The problem of finding a solution of a general homogeneous vector
equation $ H \overline{E^{p}} = 0$ is invariant under left side
multiplication by any matrix F, yielding

\begin{equation}
\label{eq:gen_homo}
F H \overline{E^{p}} = 0
\end{equation}

Of course this invariance does not hold for non-zero eigenvalues. Thus,
Equation (\ref{eq:gen_homo}) yields different solutions for different
preconditioning matrices F.  In particular for a perturbed problem
(\ref{eq:pert2}) the solutions are no longer equivalent. Thus, in case
of noise influences different starting points for the solution
strategy can be chosen.

Some of the different strategies are mentioned below:

\begin{enumerate}
\item $H_0    := \mid E1 \mid^2 * \ \overline{U} -  IF$
\item $H_{IFn} := \overline{U} - \mid E1 \mid^{-2} * \ IF $
\item $H_{IFU} := IF^{-1} * \mid E1 \mid^{-2} * \ \overline{U} - 1 $
\end{enumerate}  

In particular, a difference is expected for points where $\mid E1 \mid$
or $\mid E2 \mid$ vanishes or nearly vanishes. If $\mid E1 \mid$ or
$IF$ is represented in a limited discrete number space on a detector
(such as a detector with $n_{bit}$ Bit resolution per pixel) the value
of $\mid E1 \mid$ or $IF$ attributed to a pixel can become exactly
zero. The points where preconditioning is not (or not 'well') defined
can be exempted from preconditioning. This strategy will be applied
for all numerical results in this paper.

Of course all different formulations are equivalent for the exact
problem ('no noise').

\section{Numerical results for the beam expander HOLOCAM}
\label{num_res}

In a first set of numerical experiments, the error amplification factor
$A_{error}$ will be calculated for the expansion ratios $\gamma = 2$
and $\gamma = 4$.  Two solution strategies will be used: either a
solution without preconditioning ('pure') or a solution with IFU
preconditioning. In all cases, the eigenvector to the lowest
eigenvalue will be determined. The results are calculated for a
14-Bit detector with 1.000 pixels. Hence, quantization is taken as a noise
source.

\begin{figure}
\begin{center}
\includegraphics[scale=0.4]{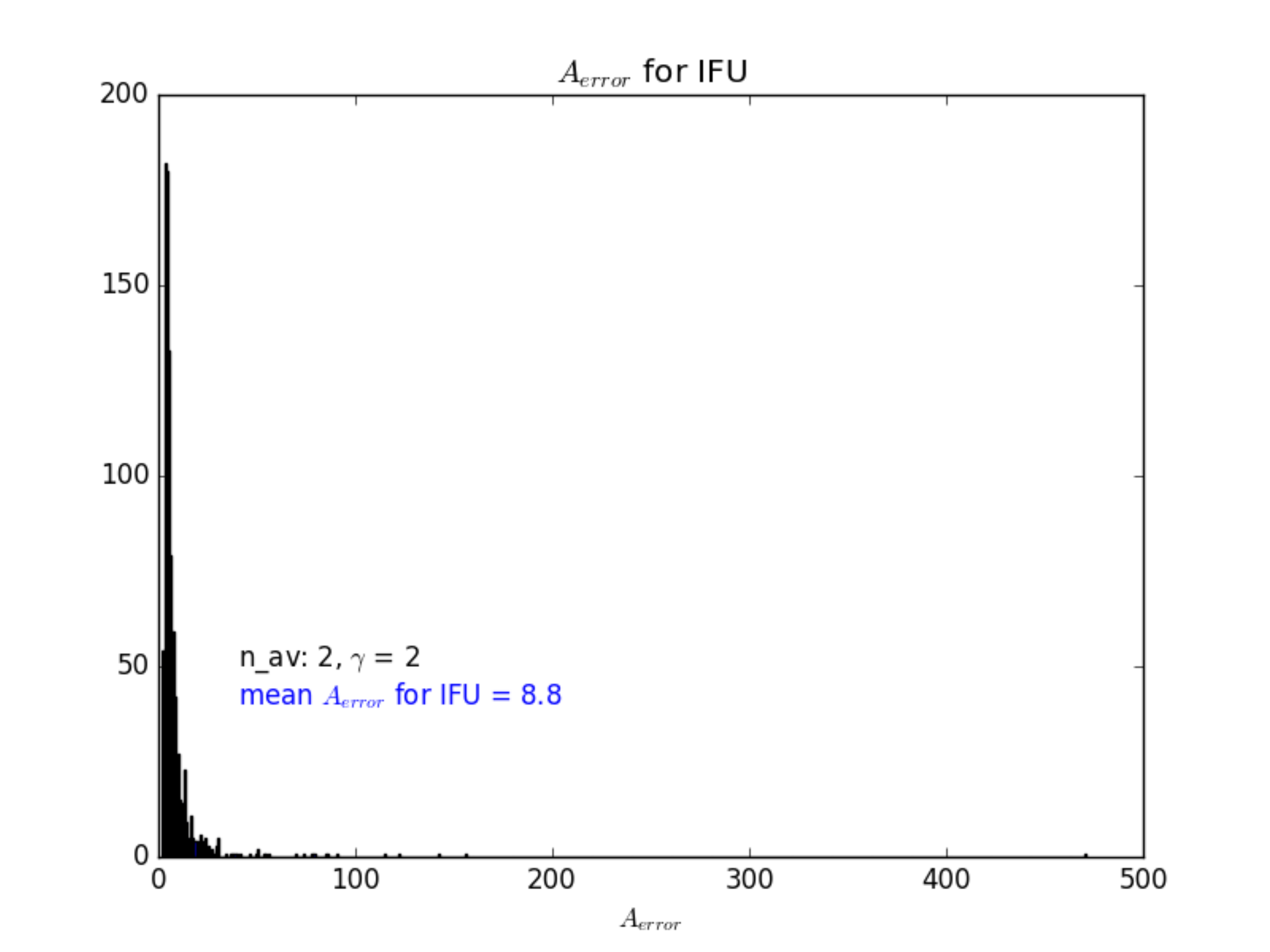}
\caption{\label{fig_hist_10f} Histogram (1.000 runs) of the error
  amplification factor $A_{error}$ for the beam expander setup with
  IFU preconditioning, Uf, $n_{av}=2$, $\gamma = 2$. The noise source is
  quantization noise of a 14-Bit detector. }
\end{center}
\end{figure}

Every data point in the following figures is actually an average over
1.000 runs for different random input fields. Figure
(\ref{fig_hist_10f}) shows a histogram of the different results.

\begin{figure}
\begin{center}
\includegraphics[scale=0.4]{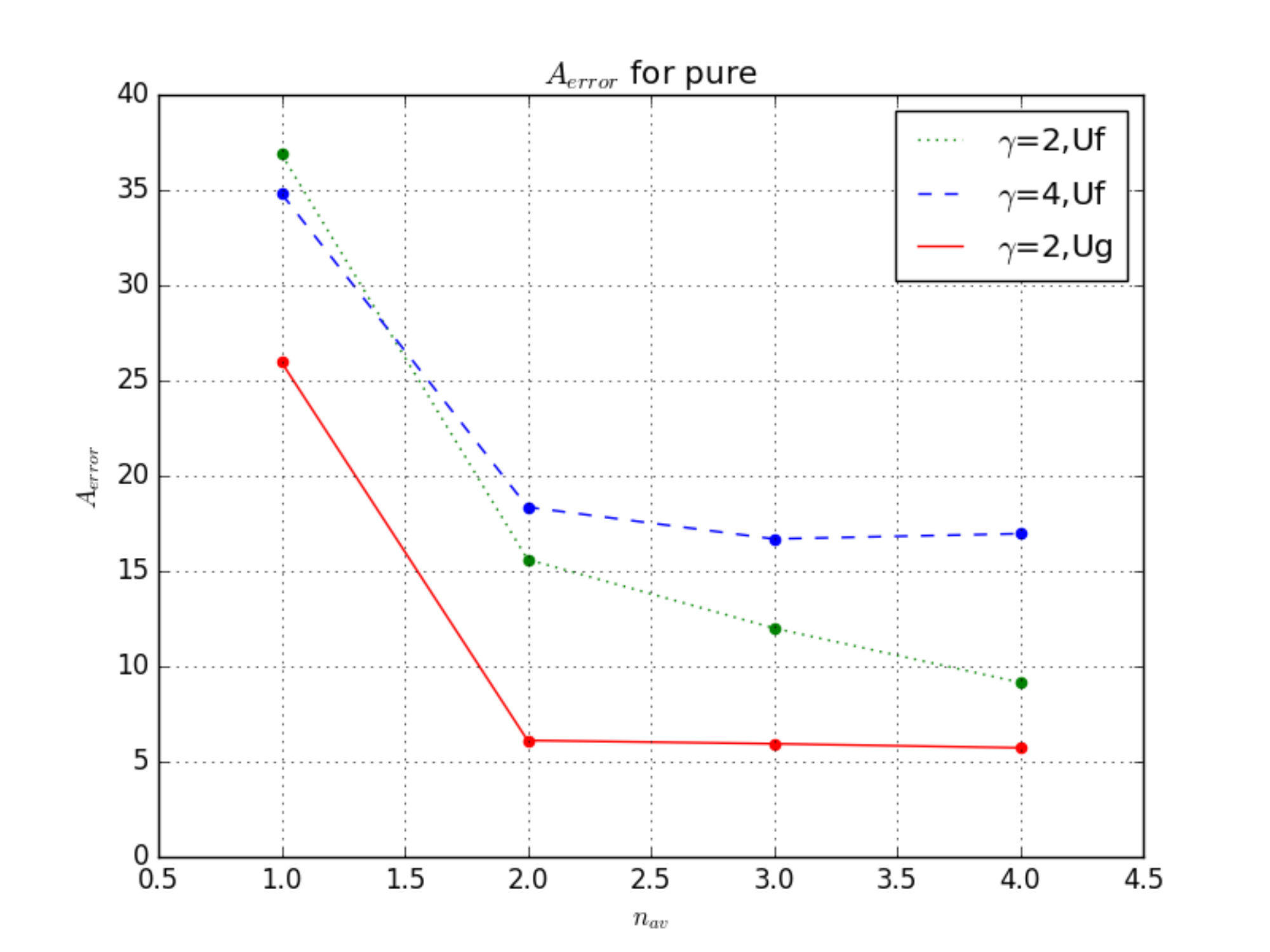}
\caption{\label{fig_pure_keyval_10f} The error amplification factor
  $A_{error}$ for the beam expander setup, without preconditioning
  (i.e. 'pure'). The noise source is quantization noise of a 14-Bit
  detector. $n_{av}$ is the number of pix over which was
  averaged. $\gamma$ is the expansion ratio of the beam expander in
  Figure \ref{bex_setup}. $U_x$, $U_f$ specifies the propagation
  matrix.}
\end{center}
\end{figure}
 
\begin{figure}
\begin{center}
\includegraphics[scale=0.4]{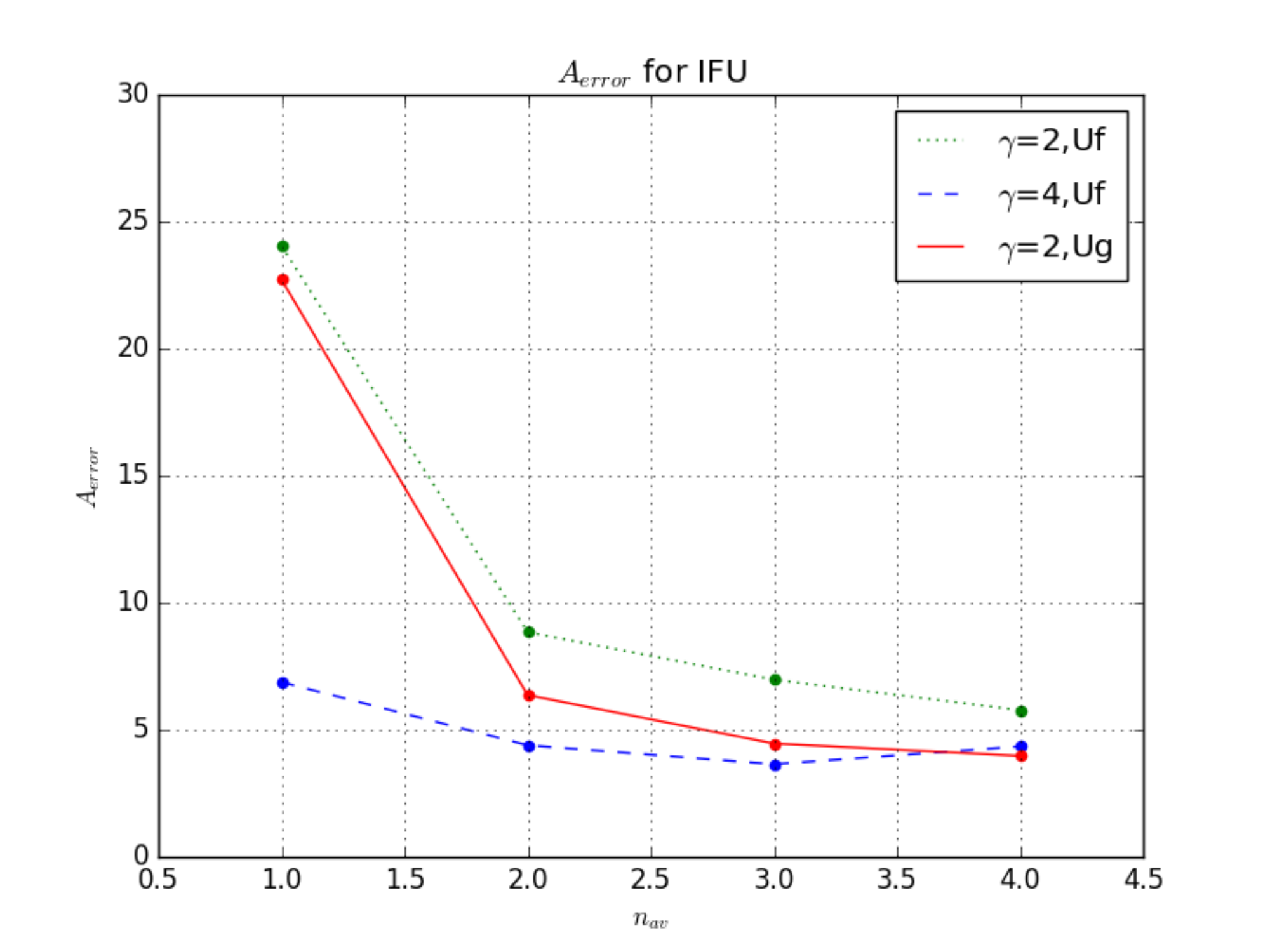}
\caption{\label{fig_IFU_keyval_10f} The error amplification factor
  $A_{error}$ for the beam expander setup using IFU preconditioning
  (i.e. 'IFU'). The noise source is quantization noise of a 14-Bit
  detector. $n_{av}$ is the number of pix over which was
  averaged. $\gamma$ is the expansion ratio of the beam expander in
  Figure \ref{bex_setup}. $U_x$, $U_f$ specifies the propagation
  matrix.}
\end{center}
\end{figure}

It can be seen from Figure \ref{fig_pure_keyval_10f} that, as
expected, the error amplification decreases for larger averaging
values. $A_{error}$ below $n_{av} = 2$ becomes larger as expected from
the Nyquist Shannon theorem since these experimental conditions could
lead to ambiguous IF values. The results are shown for two different
formulations of U: either geometric average in U ($U_x$) or Fourier
average in U ($U_{f}$). Figures
\ref{fig_pure_keyval_10f},\ref{fig_IFU_keyval_10f} represent the results
without and with preconditioning, respectively.

The results in Figure \ref{fig_IFU_keyval_10f} show that
preconditioning can improve the solution stability for certain
applications. Moreover, a smaller error amplification is observed for
larger $\gamma$ values (as expected).

\begin{figure}
\begin{center}
\includegraphics[scale=0.4]{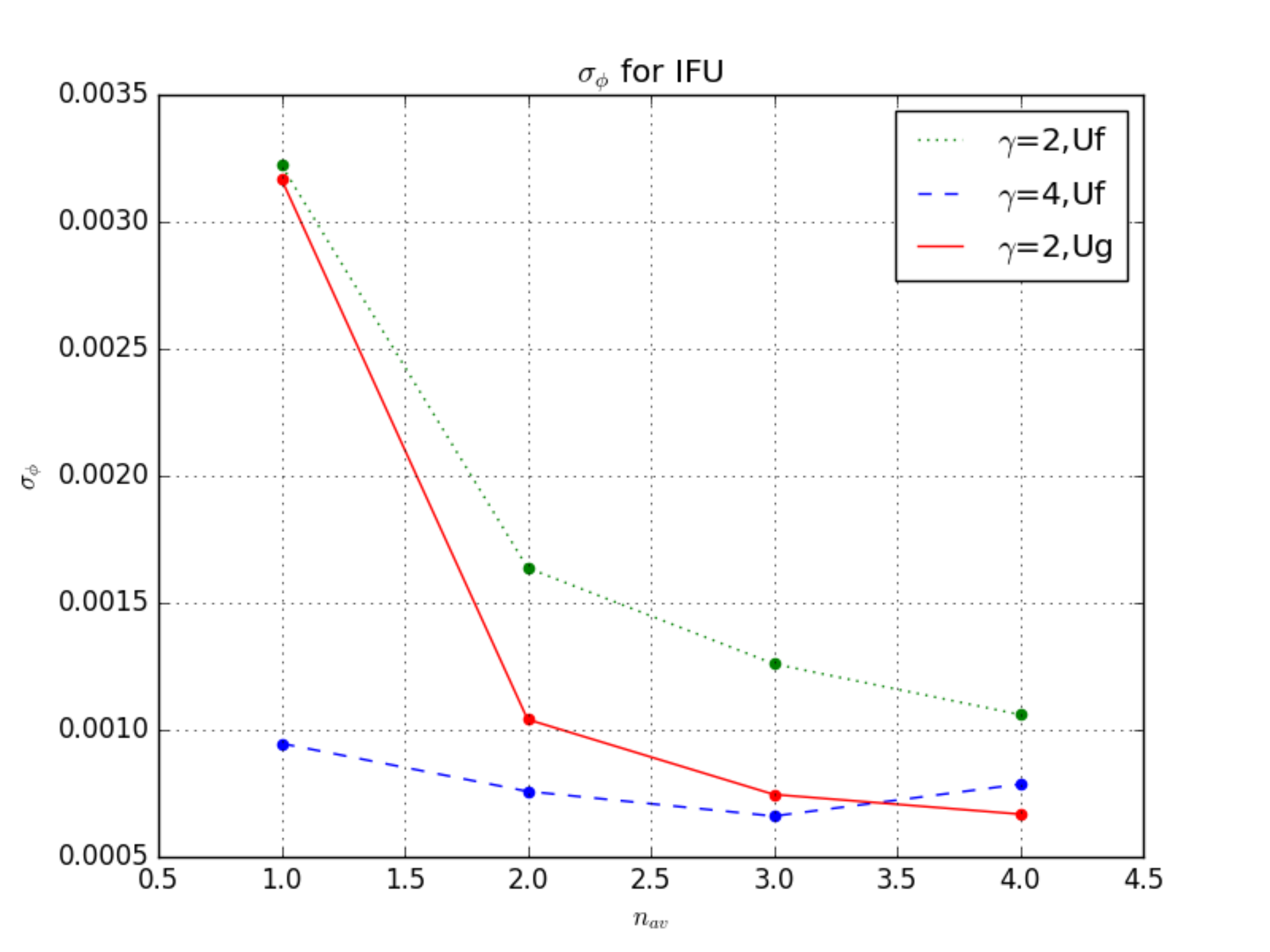}
\caption{\label{fig_IFU_sigma_10f} The $\sigma_{\phi}$ variance of the
  phase determination error for the beam expander setup, using IFU
  preconditioning (i.e. 'IFU'). The noise source is quantization noise
  of a 14-Bit detector. $n_{av}$ is the number of pix over which
  was averaged. $\gamma$ is the expansion ratio of the beam
  expander in Figure \ref{bex_setup}. $U_x$, $U_f$ specifies the
  propagation matrix.}
\end{center}
\end{figure}

The phase variance of $1e^-3$ corresponds to a factor
$A\phi_{BIT-error}$ = 16.3. The variation of the phase signal is $\pm
\pi$, hence the phase variance has to be divided by $\pi$ to get a
normalized value that can be compared to $A_{error}$. As a result, both
values $A\phi_{BIT-error}$ and $A_{error}$ are in reasonable
agreement.

Figure \ref{fig_hist_10f} shows that most runs yield values in a
narrow range about E1. Some runs lead to bigger errors. Although the
statistical weight of these erroneous data is small they are included in the
average of $A_{error}$ (8.8). $A_{error}$ = 3.5 is the factor with
the largest statistical weight.

The character of the problem and its solution is further elucidated by
the inspection of single run results. A random run with an expansion
ratio of $\gamma=2$ and no further averaging ($n_{av} = 1$) is chosen,
using $U = U_f$ .

\begin{figure}
\begin{center}
\includegraphics[scale=0.4]{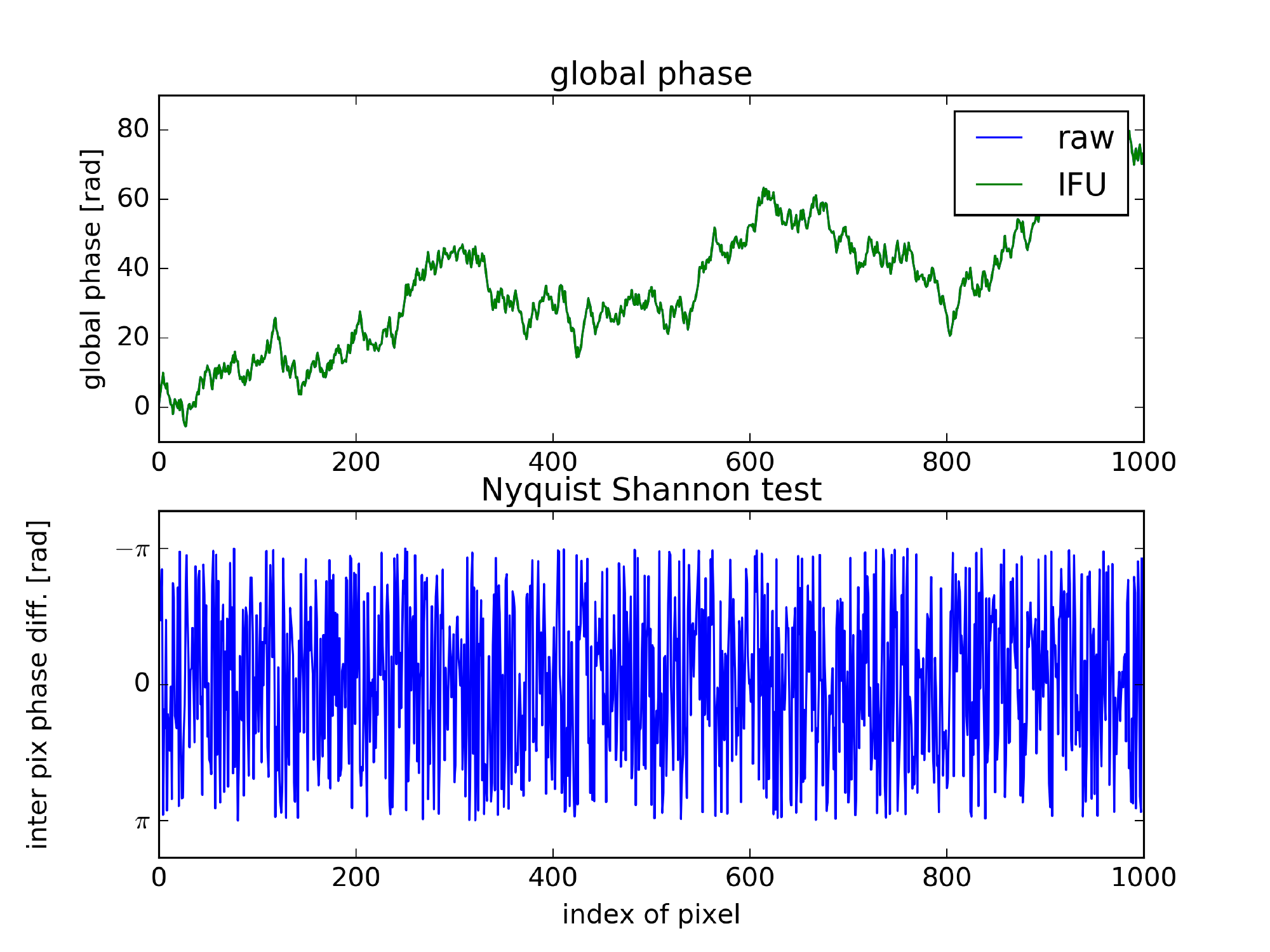}
\caption{\label{fig_glob_phase_sample}The global phase and the phase
  difference between adjacent pixels. The upper figure shows the
  global phase for the unperturbed field $E1$ ('raw' or 'original')
  and the field obtained for the perturbed problem using IFU
  preconditioning. The global phases are obtained by phase
  unwrapping. The lower figure represents the phase difference between
  adjacent pixels for the unperturbed field. The random character can be
  seen.}
\end{center}
\end{figure}

Figure \ref{fig_glob_phase_sample} shows the initial phase field for
the test run. The global phase, as obtained by phase unwrapping, fluctuates in a range of about 80
radians. The phase difference between adjacent pixels is absolutely random, i.e. random
in the range $[-\pi, +\pi]$. 

\begin{figure}
\begin{center}
\includegraphics[scale=0.4]{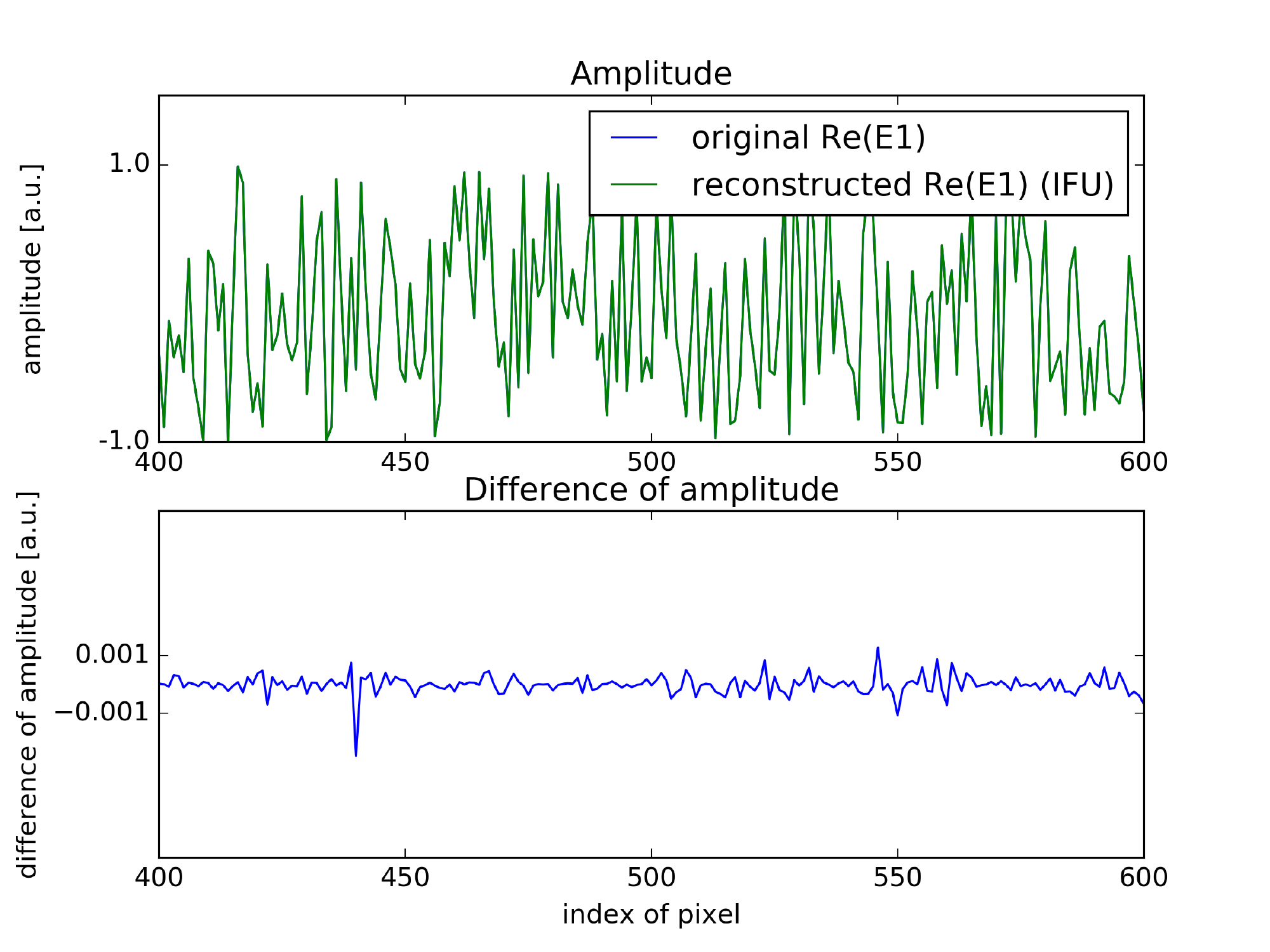}
\caption{\label{fig_e1_sample} The real part of the complex field E1
  for the unperturbed 'raw' or 'original' field and the reconstructed
  field as obtained by solving the fundamental Equation
  (\ref{eq:fund_eq}). The lower curve shows the difference between the
  raw field and the reconstructed field. It is plausible that the
  error is smaller than $1e^{-3}$ rad (for a 14 Bit detector). The
  restricted range of pixels from 400 to 600 is shown. }
\end{center}
\end{figure}

\begin{figure}
\begin{center}
\includegraphics[scale=0.4]{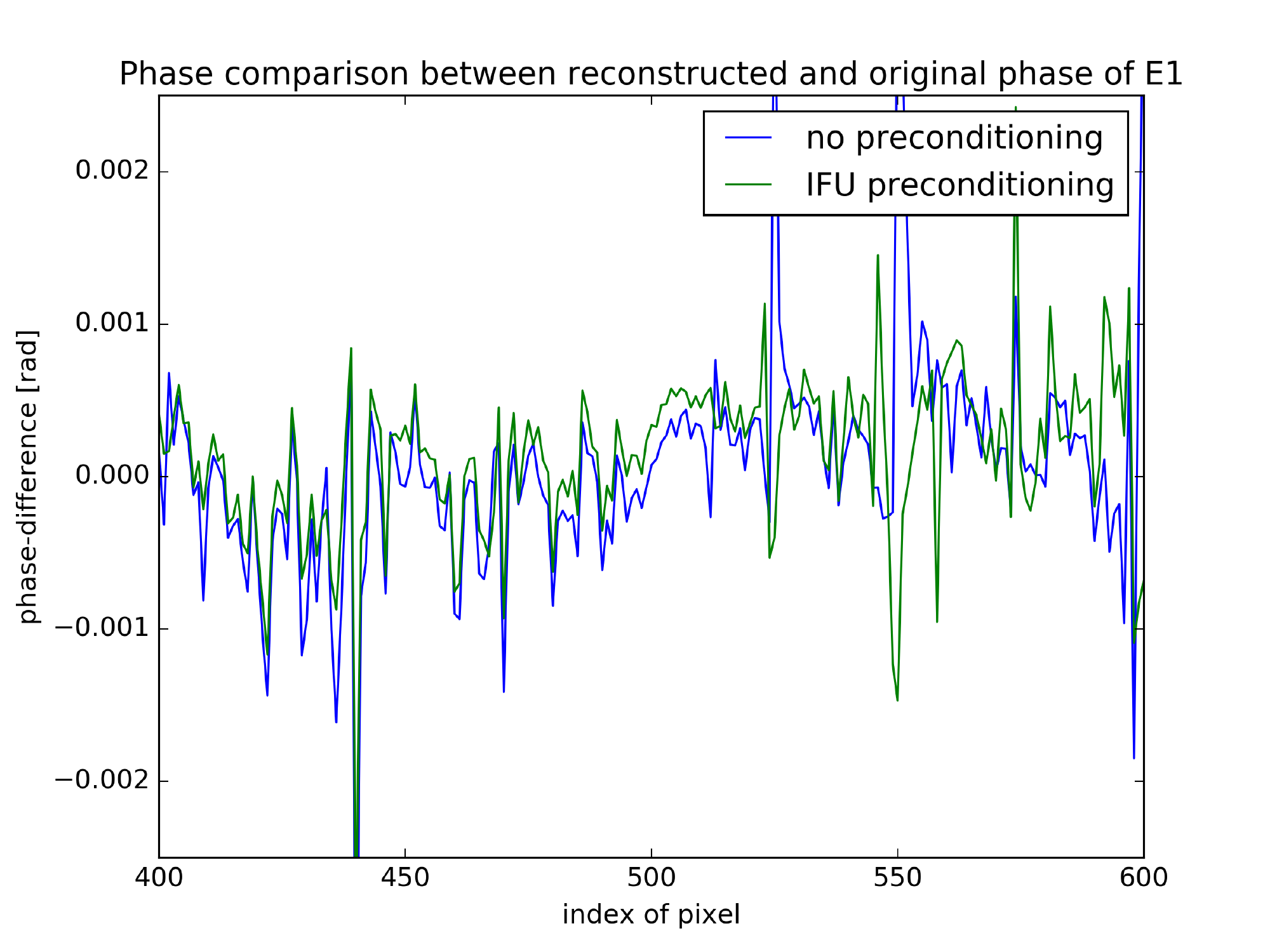}
\caption{\label{fig_phase_comp_sample} Influence of preconditioning on
  the reconstruction quality. The
  restricted range of pixels from 400 to 600 is shown.}
\end{center}
\end{figure}

Figure \ref{fig_e1_sample} depicts the real part of the 'raw' or
'original' field and the reconstructed field. For clarity, the lower
figure shows the difference between the two fields.  In this
particular case, there is little difference between 'no
preconditioning' and 'IFU preconditioning' as shown in Figure
(\ref{fig_phase_comp_sample}).

\begin{figure}
\begin{center}
\includegraphics[scale=0.4]{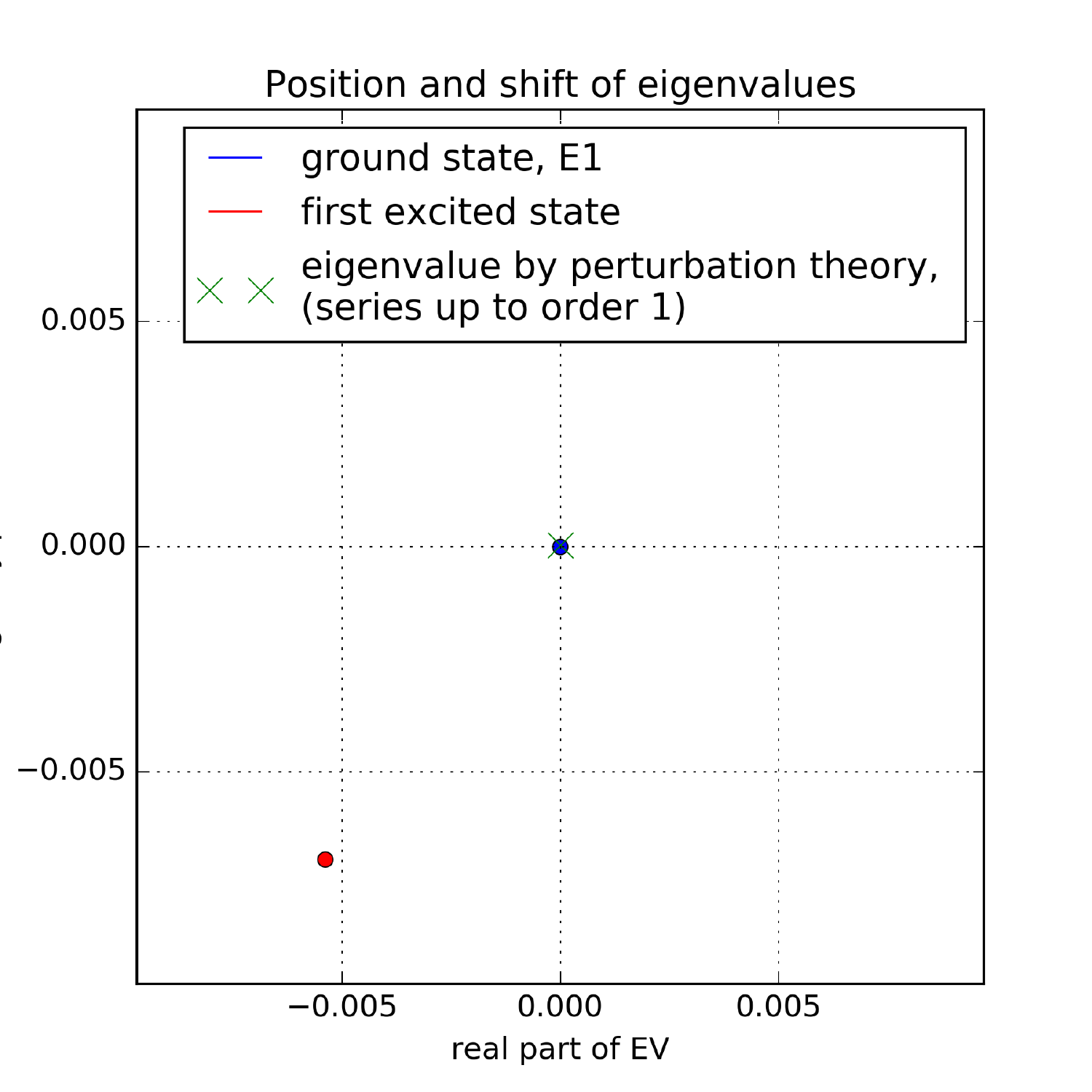}
\caption{\label{fig_eigenval_sample}Eigenvalues of the fundamental
  equation. The lowest eigenvalues of the unperturbed
  problem and the eigenvalue as obtained by perturbation theory are shown. For
  14 Bit quantization noise the perturbed eigenvalue stayed almost
  immobile at the initial location.}
\end{center}
\end{figure}

As known from perturbation theory in quantum mechanics, the influence
of perturbations on selected eigenstates comes about by the interaction
coupling to neighboring states. This arises if the eigenvalues cross as
a function of the interaction parameter ($\gamma$). Figure
\ref{fig_eigenval_sample} actually shows that such a crossing does not
occur. The eigenvalue of first order perturbation theory remains in
the neighborhood of the ground state with eigenvalue zero. 

\begin{figure}
\begin{center}
\includegraphics[scale=0.4]{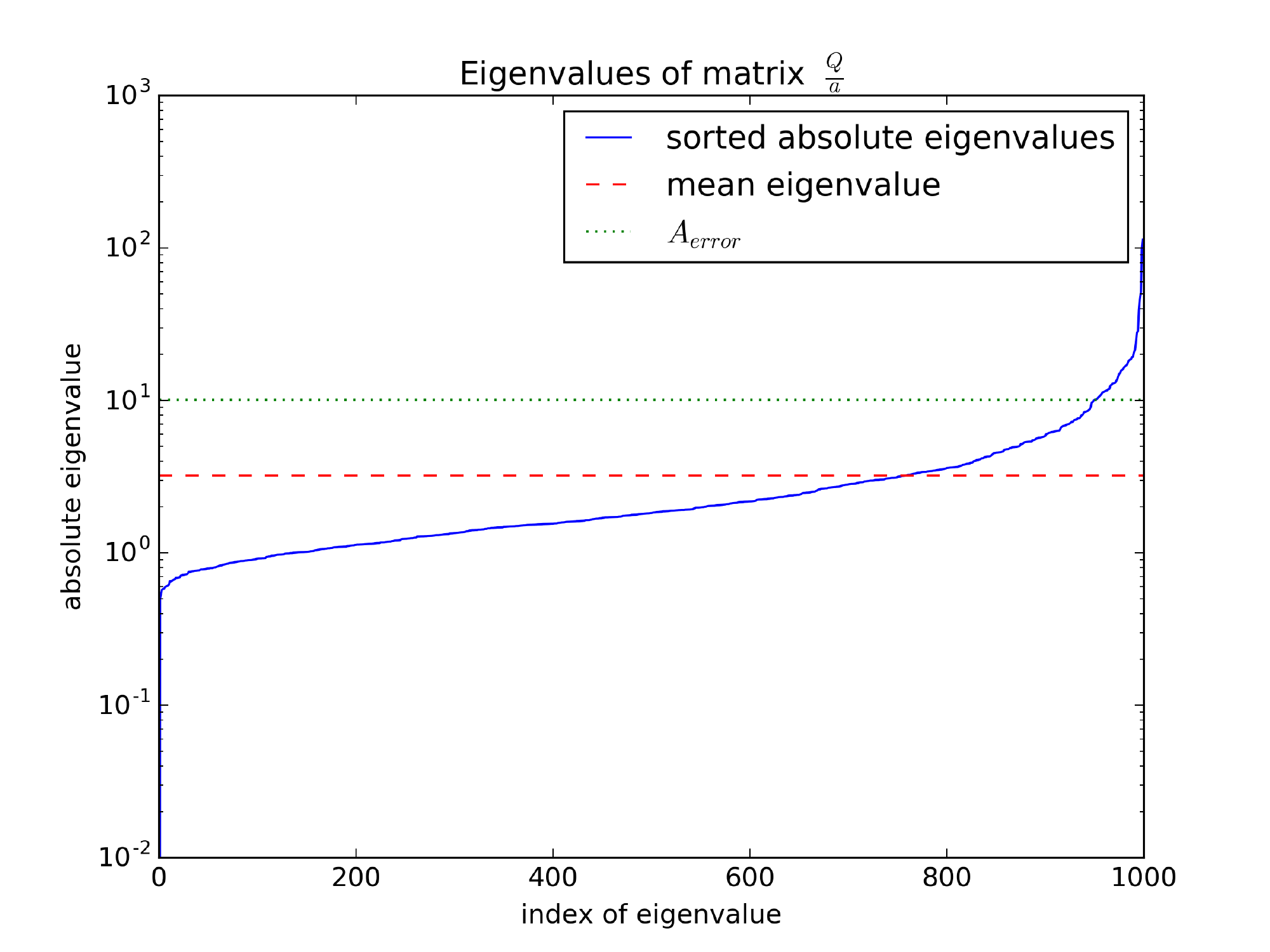}
\caption{\label{fig_eigenval_Qa_sample}Eigenvalues of the error
  propagation matrix $\frac{Q}{a}$. The eigenvalues are mainly located
  in the range 1 to 10. Besides the error propagation factor
  $A_{error}$ and the mean eigenvalue (for $n_{av}=1$, $\gamma = 2$,
  $U_f$, H = 'pure') are plotted as vertical lines.}
\end{center}
\end{figure}

Equation (\ref{eq:1st_order_compact}) gives an explicit expression for
the error propagation. As a consequence, the sensitivity is given by the spectrum
of eigenvalues of $\frac{Q}{a}$. The explicit calculation results are
shown in Figure \ref{fig_eigenval_Qa_sample}. 

\begin{figure}
\begin{center}
\includegraphics[scale=0.4]{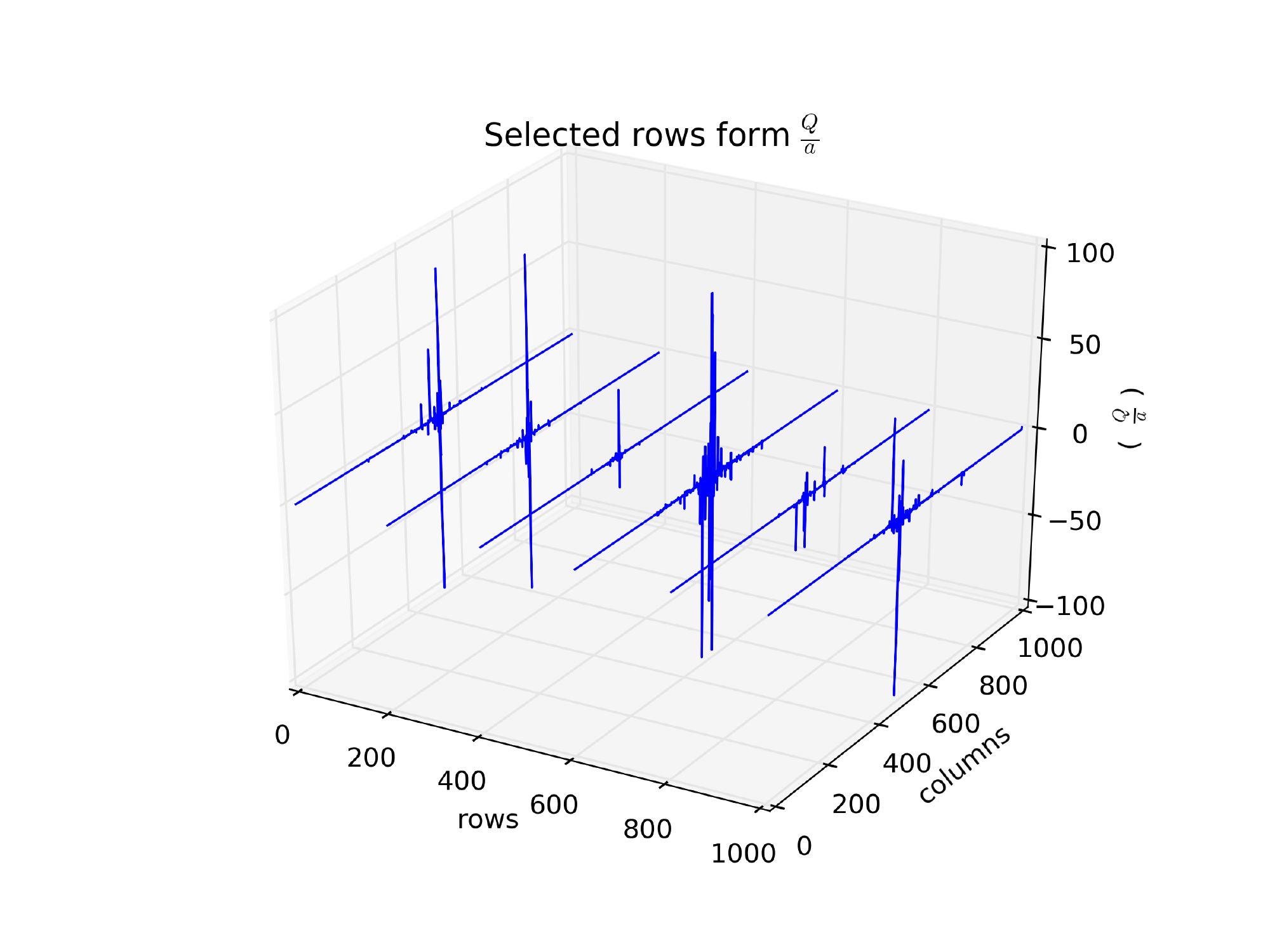}
\caption{\label{fig_rows_Qa_sample} Selected rows of the error
  propagation matrix $\frac{Q}{a}$. For selected row values all column
  values are shown.}
\end{center}
\end{figure}

The structure and explicit matrix
values of $\frac{Q}{a}$ can be further analyzed. Selected rows are shown in Figure
\ref{fig_rows_Qa_sample}. It can be seen that the matrix has a spiky
structure which explains that larger $A_{error}$ values are
rarely observed. These larger values still remain in some reasonable
neighborhood to the mean value (histogram in Figure \ref{fig_hist_10f}).

The former analysis is mainly based on a 14 Bit data acquisition. This
means that the phase of IF is represented by a 14 Bit
integer. Alternatively, the perturbation can be represented by a
gaussian noise source as described in Section \ref{sensitivity}. The
advantage of this approach is that larger values of $\sigma_{\delta
  IF}$ can be continuously probed thereby testing the extend of the
linear error propagation. This will be done for
$\sigma_{\delta IF}$ values from $1e^{-4}$ to $\sigma_{\delta IF} =
1$. The latter corresponds to an IF signal that is essentially buried
by noise of equal amplitude. No useful phase information is expected
to be found for such a noisy situation.

\begin{figure}
\begin{center}
\includegraphics[scale=0.4]{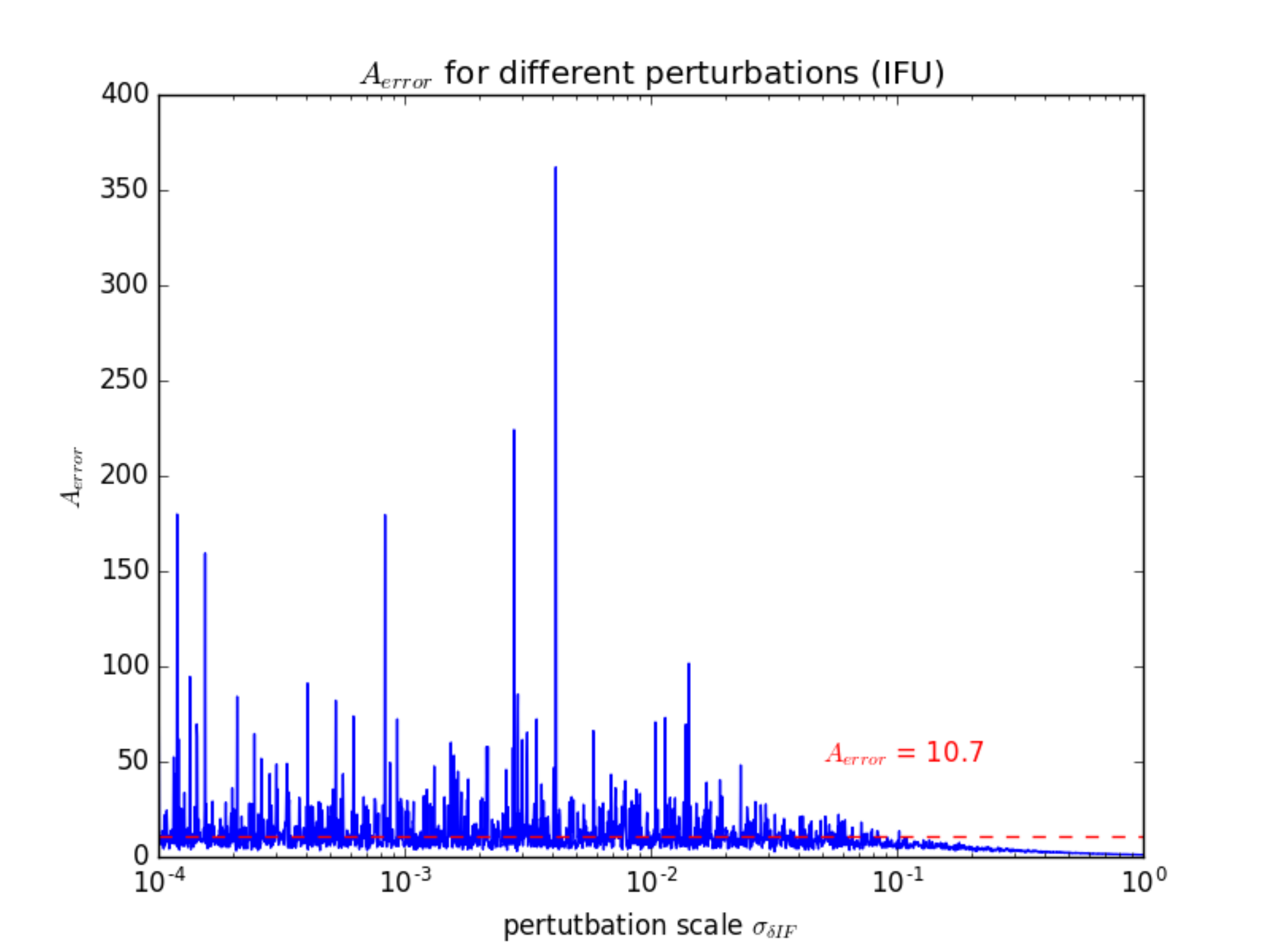}
\caption{\label{fig_A_r_res} $A_{error}$ for different perturbations
  $\sigma_{\delta IF}$. $A_{error}$ is constant over almost three
  order of magnitude. For still higher noise contributions $A_{error}$
  decreases.}
\end{center}
\end{figure}

Figure \ref{fig_A_r_res} shows an essentially linear relation between
the error in $\overline{E1^p}$ and the intensity of the noise
source. The linear factor $A_{error}$ remains almost constant up to
$\sigma_{\delta IF} = 0.1$. For higher noise levels, a sublinear
behavior is observed. For practical applications it is important not
to have a super-linear behavior. The latter would correspond to some
kind of 'breakdown' of the method which is happily not observed.

\begin{figure}
\begin{center}
\includegraphics[scale=0.4]{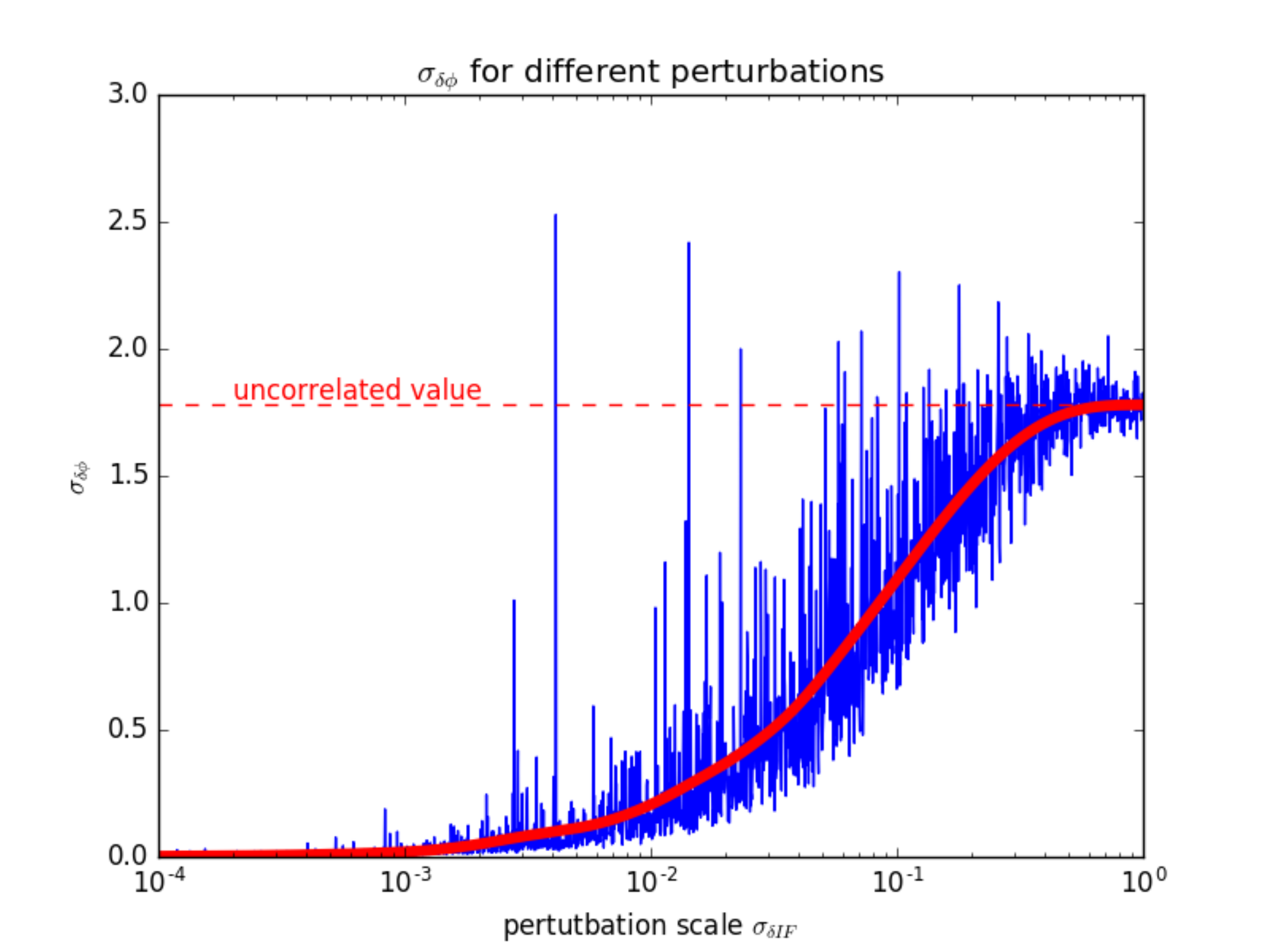}
\caption{\label{fig_phir_res} $\sigma_{\delta \phi}$ for different
  perturbations $\sigma_{\delta IF}$. The figure shows a smooth
  increase of the phase noise in $\overline{E1^p}$. The phase noise
saturates as expected at higher noise contributions.}
\end{center}
\end{figure}

Furthermore, Figure \ref{fig_phir_res} supports the interpretation of
a smooth increase in the deviation of the reconstructed field
$\overline{E1^p}$ from the unperturbed 'original' or 'raw' values. The
standard deviation $\sigma_{\delta \phi}$ expresses the degrees of
correlation. Figure \ref{fig_phir_res} shows a smooth transition from
strongly correlated (small $\sigma_{\delta \phi}$) to uncorrelated. This proves the stability of the
HOLOCAM method and is a remarkable result taking into account
difficulties in phase reconstruction~\cite{falldorf}.

\begin{figure}
\begin{center}
\includegraphics[scale=0.4]{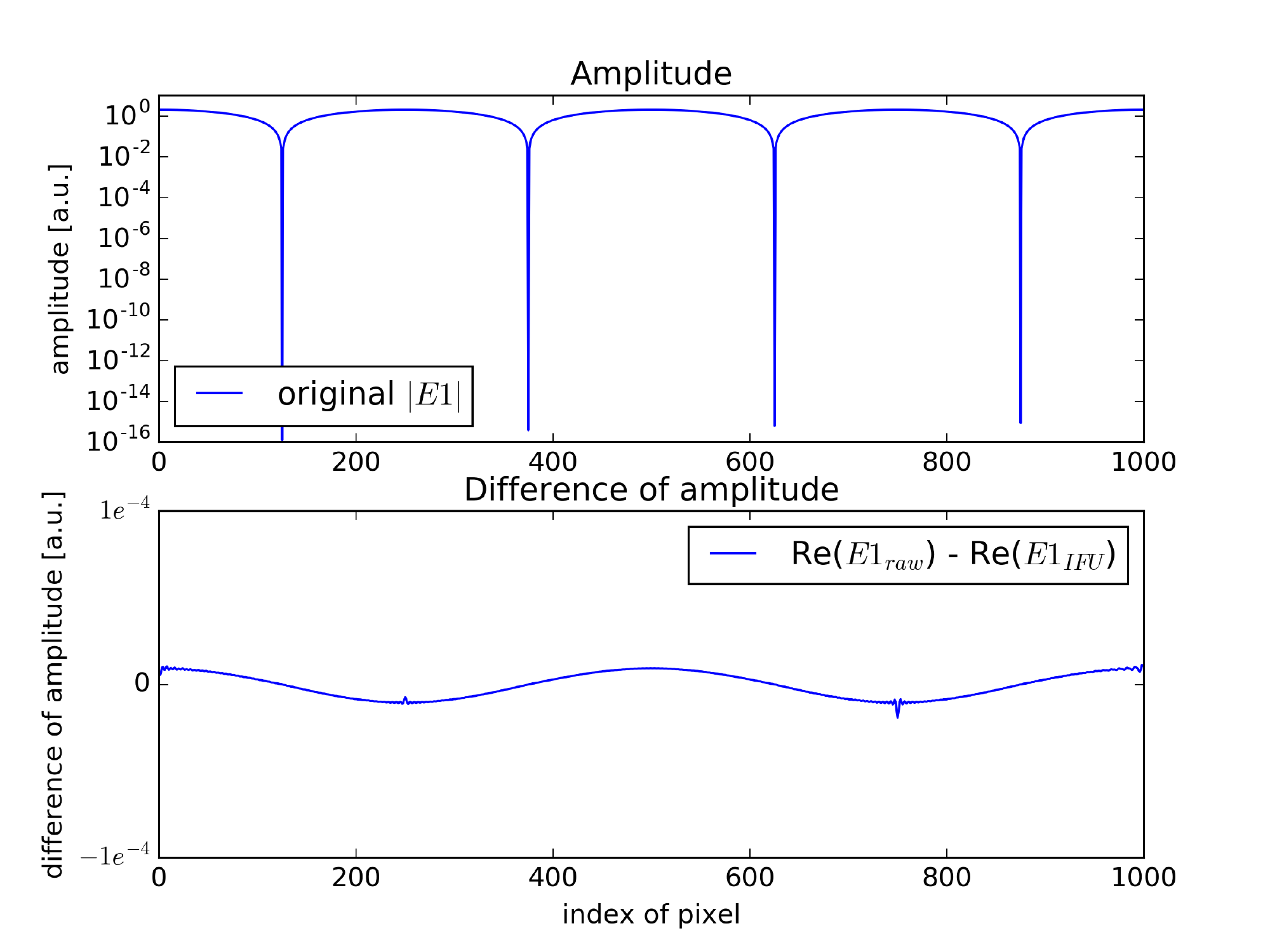}
\caption{\label{fig_sw_e1} The amplitude of the raw field $E1$ for a
  standing wave ('log plot'). As expected $|E1|$ becomes zero at some of the
  pixels. The shown field $E1$ is a standing sine wave which looks
  unfamiliar since a log scale is used for clarity. The lower figure
  shows the reconstruction quality by plotting the real difference
  between the 'raw' field and the reconstructed field.}
\end{center}
\end{figure}

Another important property of the HOLOCAM method is the resolution of
$\pi$ phase jumps. To study this property the reconstruction is tested
for $E1$ being a standing wave. A 14 Bit detector is used. Figure
\ref{fig_sw_e1} depicts the 'original' or 'raw' field as well as the
difference of the reconstructed field compared to the unperturbed
'raw' field.

\begin{figure}
\begin{center}
\includegraphics[scale=0.4]{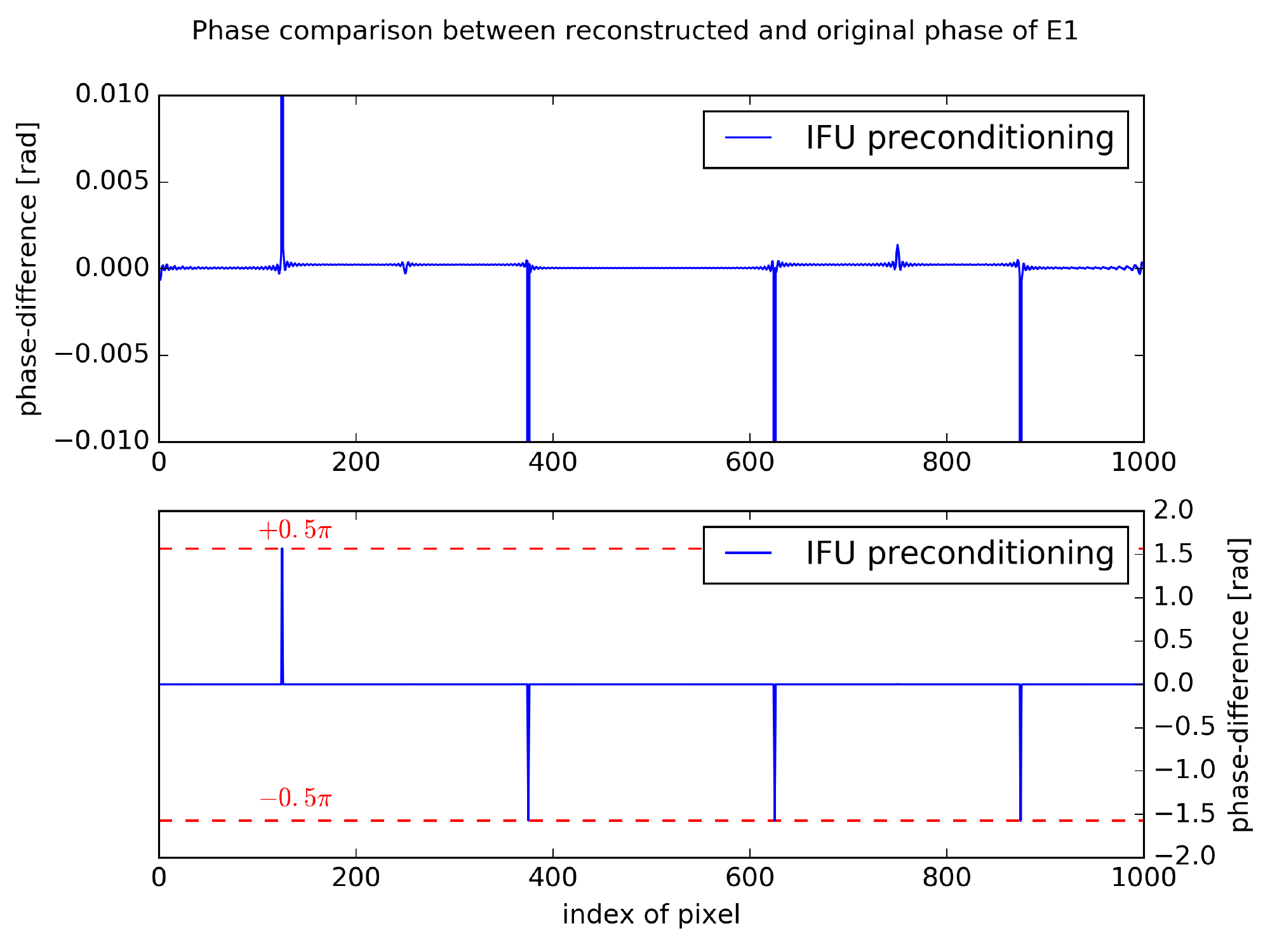}
\caption{\label{fig_sw_phase_comp} The phase difference beween the raw
  field and the reconstructed wave. The results are shown on two
  different scales.}
\end{center}
\end{figure}

The difference between the unperturbed field and the reconstructed
perturbed field is further investigated in Figure
\ref{fig_sw_phase_comp}. The upper figure shows an excellent
correspondence in the range of $1e^{-3}$ rad. Some spikes are present
which are also shown on a larger scale in the lower figure. It can be
observed that this spike is half the value of the phase jump ($0.5 \pi$). The
original 'raw' field does not have this intermediate phase
value. Therefore, this shows up as a spike. Figure \ref{fig_sw_e1}
shows that this happens at a vanishing amplitude. The difference is
not observable for the electrical field (Figure \ref{fig_sw_e1}).

\begin{figure}
\begin{center}
\includegraphics[scale=0.4]{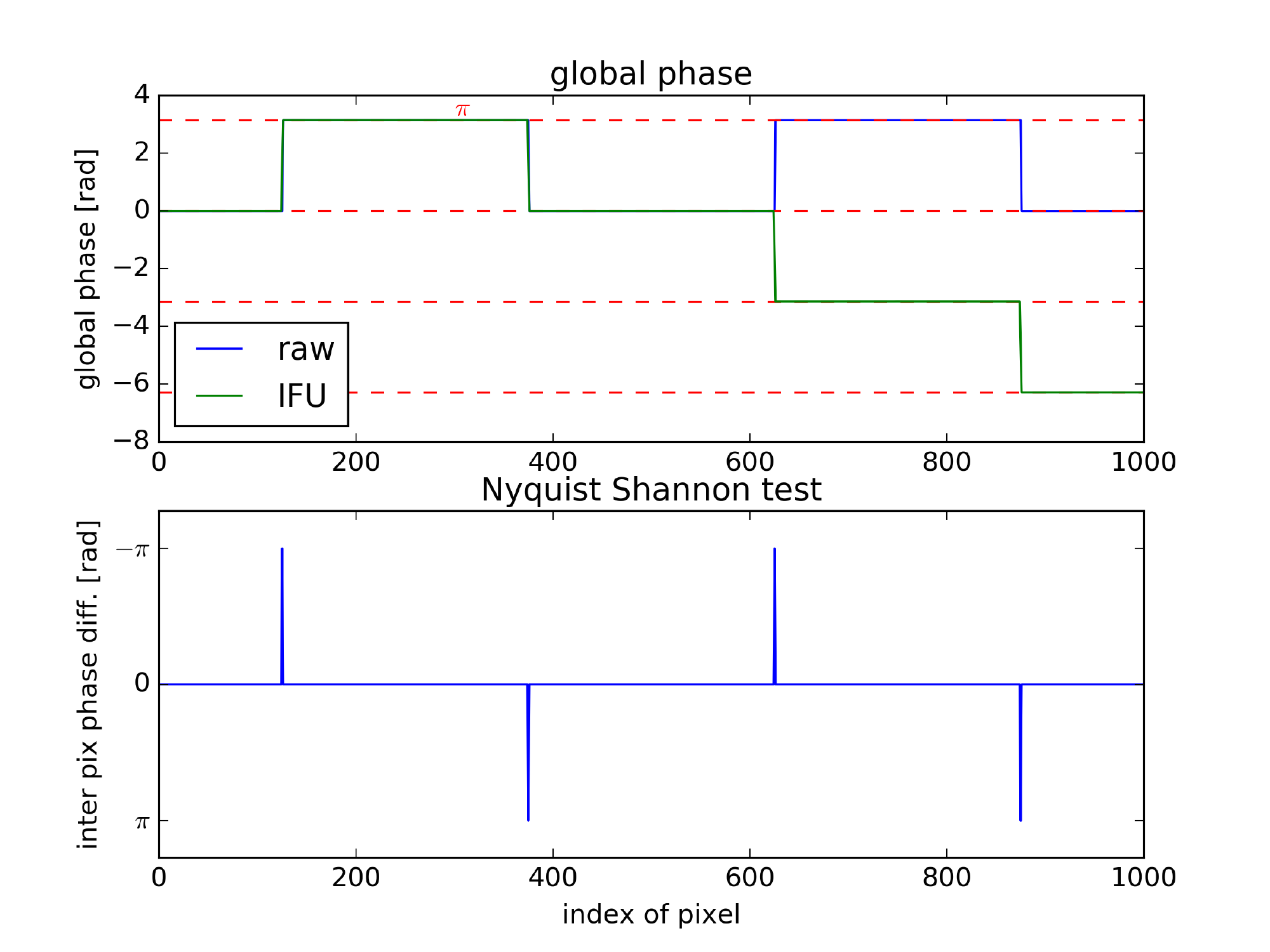}
\caption{\label{fig_sw_phase_comp2} The global phase for the raw field
  and the reconstructed wave. The right phase jump of the
  reconstructed field is separated by $2\pi$ from the 'raw' field. This
  actually corresponds to the intrinsic $2\pi$ ambiguity of the phase.}
\end{center}
\end{figure}

Figure \ref{fig_sw_phase_comp2} shows the result for the unwrapped
phase. Again an almost perfect correspondence between unperturbed
'raw' fields and reconstructed fields is found. The difference on the
right is $2\pi$ which is the intrinsic ambiguity of the phase.

\begin{figure}
\begin{center}
\includegraphics[scale=0.4]{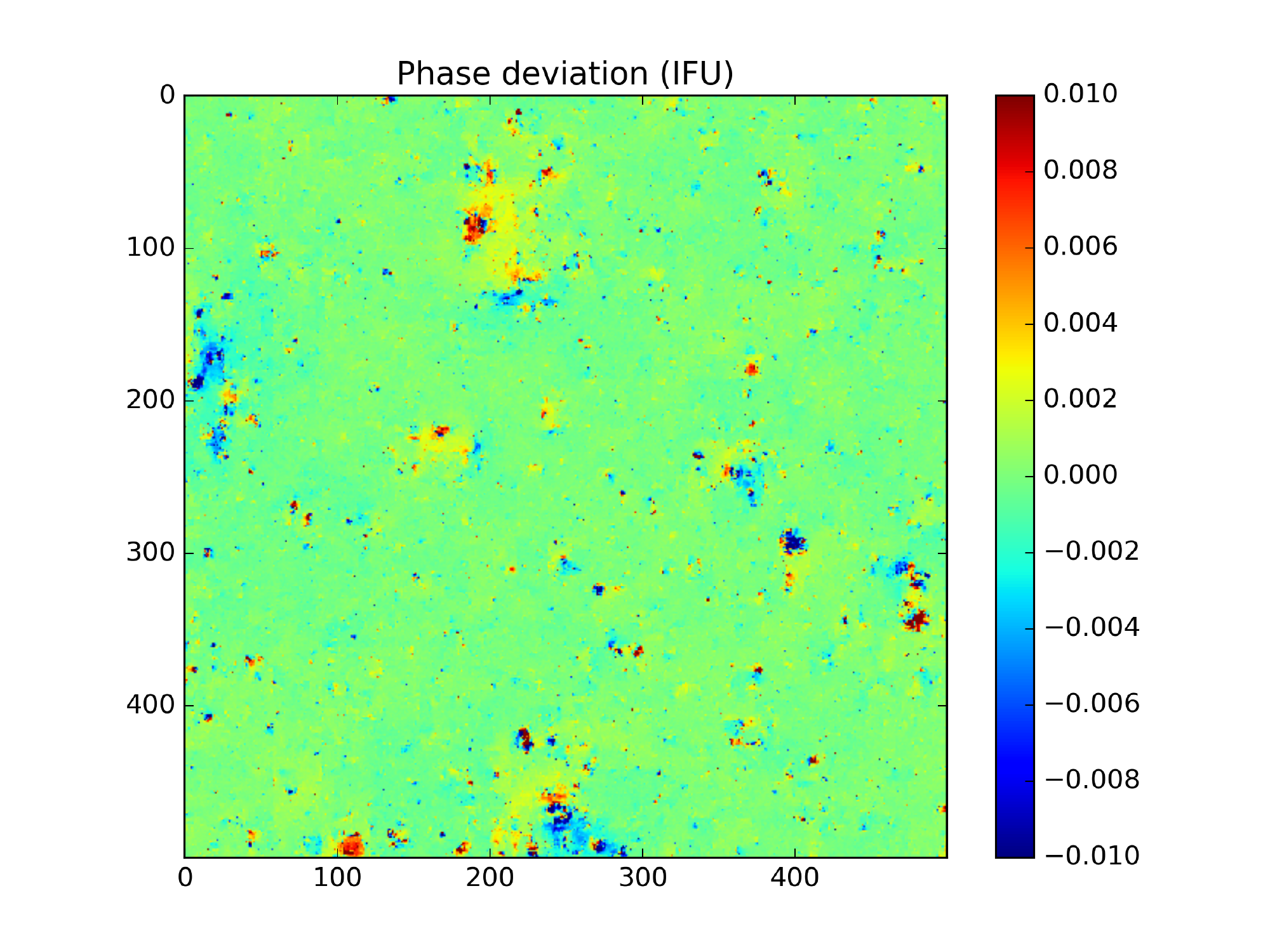}
\caption{\label{fig_phasemap_2d} 2D phasemap for a 14 Bit detector for a
  maximum random field (no correlation between adjacent pixels). Shown
  is the difference between the original 'raw' field $E1$ and the
  reconstructed field $E1^p$. The phase noise is essentially in the range
  $\pm 1e^{-3}$ as in the 1D case. The colorbar shows the phase range
  $\pm 0.01$ rad.}
\end{center}
\end{figure}

Finally, a 2D calculation is shown in Figure \ref{fig_phasemap_2d}. No
substantial difference between 1D and 2D calculations is found for the
E1 field. This might be different for the phase unwrapping step which
will not be further investigated here.

\section{The role of detector quantization}
\label{discretization}

Quantization (or detector quantization) is often the dominant noise
source. This is taken into account by choosing an integer
representation of the signal on a pixel. A 14 Bit detector means that
the signal IF is represented by two 14 Bit integers for the real and
the complex part, respectively.

\section{Finite spectral width }
\label{spectral_width}

Optical fields might have a non negligible spectral width. For this
case Equation (\ref{eq:fund_eq}) can be expressed as a frequency
dependent equation. Different frequencies separate as shown in
Equation (\ref{eq:fund_eq_finite}). The former analysis in Sections
\ref{sec_fund_equ} to \ref{num_res} also applies to this case.

\begin{equation}
\label{eq:fund_eq_finite}
IF_\omega * \overline{E1_\omega} =  \mid E1_\omega \mid^2  \ * \ \overline {U_\omega E1_\omega} 
\end{equation}

\section{The role of detector discretization}
\label{pixelation}
Discretization expresses the fact that common detectors measure the
signal on a rectangular grid of pixels. Every pixel has a finite
extension. The signal on a pixel can be assumed to be a weighted
average of the incident field intensity on the pixel.

The influence of this pixel averaging can be treated in the
framework of the fundamental Equation (\ref{eq:fund_eq}). As a result,
the HOLOCAM method can provide the functional value of the electrical
field in the center of the pixel. This a remarkable result and another
cornerstone for high precision local phase measurements.

It is recalled that the fundamental equation is formulated for a
discretization of space which must obey the Nyquist Shannon
criterion. In the following, we assume equally spaced grid points
${x_{s,t}}$ on the detector. The Fourier space ${q_{s,t}}$ is
introduced. Each Fourier vector is a 2D plane wave on the detector,
which is periodic in the boundary conditions. The set of points
${x_{s,t}}$ is finite and so is the set of Fourier vectors
${q_{s,t}}$. The maximum wavevector $q_{max}$ in ${q_{s,t}}$ is
$q_{max} = \pi / \Delta_{pix}$, $\Delta_{pix}$ being the pixel
spacing. The Nyquist Shannon criterium states that the maximum
wavevector $qE1_{max}$ found in the electrical field $E1$ must not
exceed the maximum wavevector $q_{max}$. The quantity $qE1_{max}$ is a
property of the field $E1$. It is defined independently of the
detector or pixel size. The same analysis holds for $E2$ and
$IF$. $qE2_{max}$ and $qIF_{max}$ are defined analogously. The
maximum of these three field wavevectors is called $qF_{max}$ and
$qF_{max}$ must be smaller than $q_{max}$.

The electrical fields $E1$,$E2$ as well as $IF$ are continuously
differentiable. By choosing a support large enough the field IF can be
expressed as

\begin{equation}
\label{eq:Fourier_field_e1}
IF(x)  = \sum_{q \in \{q_{ s,t} \} }^{q_{max}} \hat{IF}(q) e^{-i q x}
\end{equation}

The support is chosen such that $IF=0$ and $\nabla IF=0$ holds on the
border of the support ($\nabla$ is the gradient on the border). The
field IF therefore fulfills periodic boundary conditions.  The
summation is performed over a finite number of terms. According to Fourier
theory the Fourier expansion converges point wise to the fields. At
first glance, this applies only to the grid values ${x_{s,t}}$. But the
same question can be asked for arbitrary points x, i.e. not
necessarily at a grid point.

It is remarkable that Equation (\ref{eq:Fourier_field_e1}) holds on
every point of the continuous support, not only on the discrete
points, assumed that the function IF satisfies periodic boundary
conditions and the Nyquist Shannon condition is respected.  Choosing a
sufficiently large detector size and a sufficiently fine spaced
detector this assumption can in fact be fulfilled.

This can be proven as follows: Testing spatial points not given by the
grid ${x_{s,t}}$ is equivalent to the introduction of higher q values
in the expansion of Equation (\ref{eq:Fourier_field_e1}). The points
become regular points of the reciprocal lattice to these higher q
values. The convergence of the Fourier expansion is again point wise
but the new points are included.  Thus, the Fourier expansion has the
correct function value at the newly introduced spatial
points. According to the assumption, the Fourier amplitudes of the
newly introduced higher q modes are zero. Consequently, the functional values
for these off grid points are just given by the unchanged Equation
(\ref{eq:Fourier_field_e1}).

It is the purpose of the measurement process to determine the
amplitudes $\hat{IF}(q)$ or $IF(x)$ in Equation
(\ref{eq:Fourier_field_e1}). The Fourier transform yields

\begin{equation}
\label{eq:Fourier_analysis1}
\hat{IF}(q)  = \frac{1}{\mu_{sup} } \sum_{x_i}^{support} IF(x_i) e^{i q x_i}
\end{equation}

Here and in the following index $i$ denotes an index tupel $i=\{ s,t \}$. 
$\mu_{sup} $ is some normalization constant depending on the support
of the Fourier expansion. The difficulty with this formula is that
$IF(x_i)$ is actually unknown.

Only the pixel averaged quantities are known:  

\begin{equation}
\label{eq:Fourier_analysis2}
<IF>(x_i)  =  \int_{A_i} IF(x) dx
\end{equation}

$A_i$ being the surface of pixel $x_i$.

The quantities actually measured are $<IF>(x_i)$ and $\hat{<IF>}(q)$, with

\begin{equation}
\label{eq:Fourier_analysis3}
<IF>(x_i) =: \frac{1}{\mu_{sup} } \sum_{x_i}^{support} \hat{<IF>}(q)  e^{- i q x_i}
\end{equation}

Inserting Equation (\ref{eq:Fourier_field_e1}) in Equation
(\ref{eq:Fourier_analysis2}) yields

\begin{equation}
\label{eq:Fourier_analysis4}
<IF>(x_i)  = \sum_q^{q_{max}} \hat{IF}(q) \int_{A_i} e^{-i q x} dx
\end{equation}

which can be inverted yielding $\hat{IF}(q)$. Using the inverse
Fourier transform yields the value sought ($IF(x_i)$). Hence,
the problem of finding the non-averaged IF values on a grid can be
solved in principle.

To give an example $\int_{A_i} e^{-i q x} dx$ is evaluated with an
$A_i$-size of $\Delta_{pix}$. It yields

\begin{equation}
\label{eq:Fourier_analysis5}
\int_{A_i} e^{-i q x} dx = a \ e^{-i q x_i} \ sinc(q/q_{max}) 
\end{equation}

a being some normalization constant.  Equation
(\ref{eq:Fourier_analysis4}) can be rewritten as

\begin{equation}
\label{eq:Fourier_analysis6}
<IF>(x_i)  = a \ \sum_q^{q_{max}} \hat{IF}(q) sinc(q/q_{max}) e^{-i q x_i} \
\end{equation}

This is the Fourier expansion for the averaged values. The expansion
coefficients are unique.  As a result, comparing coefficients with Equation
(\ref{eq:Fourier_analysis3}) yields

\begin{equation}
\label{eq:Fourier_analysis7}
\hat{<IF>}(q) = \hat{IF}(q) sinc(q/q_{max}) 
\end{equation}

As expected, the non averaged quantities $\hat{IF}(q)$ are determined by
the averaged quantities $<\hat{IF}>(q)$. The relationship is a
multiplication in q-space. The inverse function is needed leading
from the averaged values to the non averaged values. This corresponds
to a folding in x-space. The kernel for this folding in real space is
called K.  K must be folded with the measured averaged data to get the
point data in demand in the center of the pixel.

\begin{equation}
\label{eq:Fourier_folding}
K(x) = K_0 \int_{|q| < q_{max}} \frac{1}{sinc(q/q_{max})} e^{-i q x} dq 
\end{equation}

The sum $\sum$ has been replaced by an integral,  $K_0$ being a
normalization constant, $q < q_{max}$. Expanding the 1/sinc function
in terms of $q / q_{max}$ yields the following approximation:

\begin{equation}
\label{eq:Fourier_folding2}
K(x) \approx  K_0 \int_{q_{max}}^{q_{max}} [ 1 + (q/q_{max})^2/ 3!  ] e^{i q x} dq 
\end{equation}

or introducing $n_g = \frac{x}{ \Delta_{pix}} $, $n_g$ being the
separation in units of $\Delta_{pix}$ ('g' stands for grating).

\begin{equation}
\label{eq:Fourier_folding3}
K(n_g { \Delta_{pix}} ) \approx  K_0 \int_{-1}^{1} [ 1 + y^2/ 3!  ] e^{i \pi y n_g} dy 
\end{equation}

\begin{figure}
\begin{center}
\includegraphics[scale=0.4]{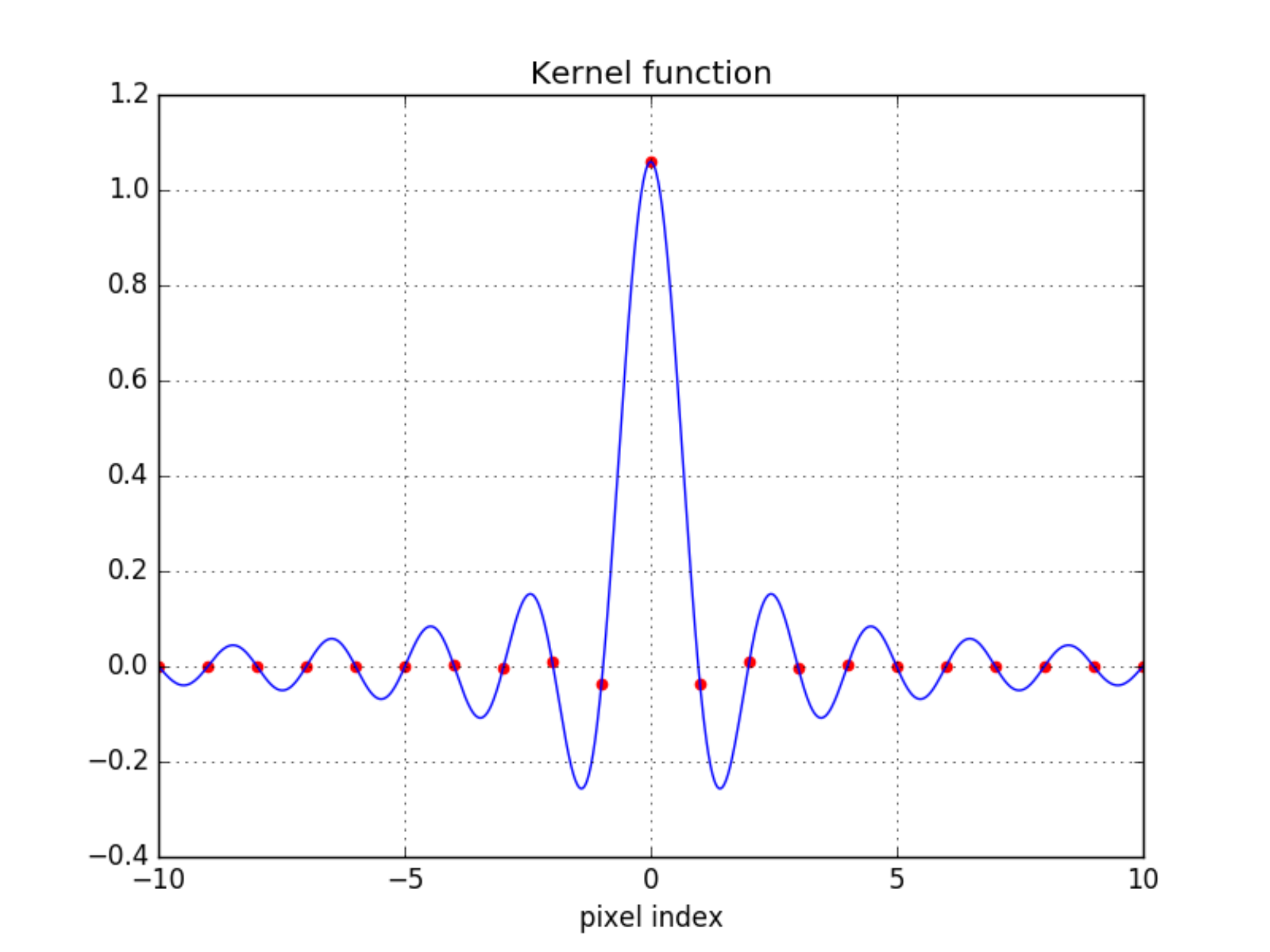}
\caption{\label{fig_kernel}Kernel for the inverse filtering from averaged pixel values to point values.}
\end{center}
\end{figure}

The integral can be calculated analytically.  The kernel function is
plotted in Figure (\ref{fig_kernel}). For the inverse folding
operation only integer values $n_g$ are needed.

\begin{equation}
\label{eq:Fourier_folding4}
\sum_i <IF>(x_i) K (x_j-x_i) = IF(x_j)
\end{equation}

Thus, Equation (\ref{eq:Fourier_folding4}) calculates, as intended, the
point values of IF out of the measured averages $<IF>(x_i)$. The
averaging effect of the detector can be treated correctly in Equation
\ref{eq:fund_eq}. This is another important cornerstone for high
precision field measurements.

\section{A preview on entangled light detection}
\label{qd_optics}

The HOLOCAM detection principle is based on space resolved
correlations. This can also be used to analyze non classical light
properties. The incoming quantum mechanical light state is called
$\Psi$. To be specific a non-classical two photon state (or Bi-photon
state) is assumed. This includes two photon number states and two
photon coherent vacuum states (squeezed vacuum states). In the
Heisenberg picture these states have a double frequency dependence of
the generating operator. The squeezing operator is given
by~\cite{loudon}:

\begin{equation}
\label{eq:squeeing-op}
\hat{S_0}(\zeta_s) = exp ( {0.5 ( \  \bar{\zeta_s}  \hat{a}_{s}^2 + \zeta_s  (\hat{a}_{s}^\dagger)^2   )} )
\end{equation}

(s is some arbitrary pixel index, 0 expresses the time independence)

In the Heisenberg picture the operator time dependence becomes:

\begin{equation}
\label{eq:squeeing-time}
\hat{S(t)}(\zeta) = \hat{S_0}(\ \zeta \ e^{2 i \omega t})
\end{equation}

This time dependence corresponds to a
doubled k vector of the Bi-photon in the Schr\"odinger picture. The Bi-photon $\Psi$ has a 2
fold space dependence $\Psi ( x_2 , x_1) $, $x_1$,$x_2$
being the coordinate of the first and the second photon,
respectively. According to Bose statistics the function is symmetric
upon permutation of the ordering. On the light path the phase changes
as $\Psi \rightarrow \Psi e ^{2 i k \delta l}$.

According to quantum optics~\cite{loudon} this light state is split
into three components by a beam splitter. Light states where both
wave parts follow path 1 or path 2 and a third option which only
exists in quantum mechanics where 1 photon passes by path 1 and 1
photon passes by path 2. These three wave functions are called
$\Psi_{20} $, $\Psi_{02} $, $\Psi_{11} $. The experimental situation
discussed corresponds to the setup in Figure \ref{bex_setup}.

On the detector, a superposition of these three wave functions is
formed. This step of superposition must be done carefully since
further wave components are generated on the second beam splitter. The
effects on beam splitter 1 and beam splitter 2 are quite similar. In
the experiment described here, only the states where both photons
reach the detector are of interest. Other states should be ideally
rejected.

Different solutions exist to this problem. It is the conceptually
simplest approach to limit oneself to two photon number states and to
use a two photon detector~\cite{loudon}. Thus, this is chosen here. As
a consequence, Bi-photon splitting on the second beam splitter will be
neglected and the following interference signal QIN is measured:

\begin{equation}
\label{eq:qd-interference}
QIN_s = \bra{ \Psi_{20} + \Psi_{02} + \Psi_{11}  } N(x_s) \ket{ \Psi_{20} + \Psi_{02} + \Psi_{11}}
\end{equation}

$N(x_s)$ is the 2 photon number operator ($ (\hat{a}_{s}^\dagger)^2
(\hat{a}_{s})^2 $) representing the detector pixel $x_s$. To some
extent QIN replaces the previously introduced quantity IN (Section
\ref{sec_fund_equ}) although differences exist.

As usual in interference experiments, the double sum in Equation
(\ref{eq:qd-interference}) can be multiplied out yielding intensity
and interference terms.

There is only one term (and its conjugate complex) which has a
Bi-photon path length dependence such as in Equation
(\ref{eq:squeeing-time}).

\begin{equation}
\label{eq:qd-interference2}
QIN_s = \bra{ \Psi_{20}} N \ket{ \Psi_{02} } e ^{2 i k \delta l} + c.c. + 'other terms'
\end{equation}

'other terms' designates terms which do not have the 'double k'
dependence on path variations $\delta l$ in the interferometer.

Hence, using a phase resolving method, such as phase shifting,
$\bra{  \Psi_{20}} N (x_s) \ket{ \Psi_{02} }$ and furthermore
$\bra{\Psi_{20}} N (x_s) \ket{ \Psi_{20} }$ and
$\bra{\Psi_{02}} N (x_s) \ket{ \Psi_{02} }$ can be measured.

The space index in $N (x_s)$ indicates that every pixel is
measured. The methods have already been used for Equation
(\ref{eq:basic_IF4}).

Knowing the detector response~\cite{loudon} yields the quantity

\begin{equation}
\label{eq:biif_2}
QIF_{s,s} = \overline{\Psi2_{s,s}} \Psi1_{s,s} 
\end{equation}

and similarly the quantities $|\Psi2_{s,s}|^2$ and $|\Psi1_{s,s}|^2$.
The short notation $\Psi1_{s,s}$, $\Psi2_{s,s}$ has been introduced
for $\Psi_{20}(x_s,x_s)$, $\Psi_{02}(x_s,x_s)$
respectively. The equation holds for equal values in the x-entry slots
of $\Psi_{20}(x_s,x_s)$. The double index with identical
values indicates that the quantity denotes properties of the Bi-photon
field for one and the same index.

The measuring device comprises a first beam splitter, two arms which
are again described by some single mode propagation matrix U (Section
\ref{sec_fund_equ}). As in Section \ref{ref_BEX} an appropriate
interferometer with a point mapping U is assumed (Figure
\ref{bex_setup}). Following the rules of Section \ref{ref_BEX} a
mapping of the Bi-photon in branch 1 called $\Psi1_{s,t}$ to branch 2
can be derived.

\begin{equation}
\label{eq:biphoton_U}
\Psi2_{s,t} = \sum_{s,t} U^\diamond_{s,t,m,n} \Psi1_{m,n} = U^\diamond(\Psi1) 
\end{equation}

In general, this mapping is a linear function of the
array of indices $(s,t),(m,n)$. In the case of a point
mapping though $U^\diamond$ maps diagonal (s,s) matrix elements on diagonal
(m,m) elements. The mapping is essentially 'geometric' and there is no
smearing out by diffraction.

\begin{equation}
\label{eq:biphoton_U2}
\Psi2_{s,s} = \sum_{s} U^\diamond_{s,s,m,m} \Psi1_{m,m} = \sum_{s} U_{s,m}^2 \Psi1_{m,m}
\end{equation}

$U_{s,m}^2$ denotes the square of the U matrix element in Equation
\ref{eq:basic_U}. Thus, this is a simple rule to deduce $U^\diamond$
from $U$.

An equation similar to Equation (\ref{eq:if_1}) holds

\begin{equation}
\label{eq:biif_1}
QIF_{s,s} = \overline{\Psi2_{s,s}} \Psi1_{s,s} = \Psi1_{s,s} \sum_{t} \overline {U_{s,s,t,t}^\diamond \Psi1_{t,t}} 
\end{equation}

The solution is identical to Equation (\ref{eq:if_1}). $\Psi1_{s,s}$
denotes the complex field value of the Bi-photon state on index $(s,s)$,
s denoting for instance a pixel on the detector.

It should be mentioned that $\bra{ \Psi_{11}} N \ket{ \Psi_{02} } e ^{i k
  \delta l}$ and $\bra{ \Psi_{11}} N \ket{ \Psi_{20} } e ^{i k \delta
  l}$ potentially allow to measure 'off-diagonal' (s,t) states of
$\Psi1(s,t)$. The measured quantities are of the type $\Psi_{20}(s,s)
\Psi_{11}(s,s)$, $\Psi_{02}(s,s)\Psi_{11}(s,s)$. For this step $\Psi_{20}(s,s) $ and
$\Psi_{02}(s,s) $ must be known from the first procedure. The involved
equations can be directly solved yielding $\Psi_{11}(s,s) $.
Using the properties of U and the mapping of $\Psi_{11}(s,s) $
similar to Equation (\ref{eq:biphoton_U2}) a set of
off-diagonal elements of $\Psi1 ( s , t) $ can be
measured. By shearing the two paths of the interferometer other index
combinations can be measured.

This shows that a careful combination of HOLOCAM and shearing
methodology allows us to measure the hole set of general Bi-photon
parameters $\Psi1 (x_2 , x_1) $.

The Bi-photon HOLOCAM should be useful for super resolution microscopy
and cutting edge metrology. The information is different from homodyne
detection. This can be immediately concluded from the fact that the Bi-photon
states (such as the Bi-photon number states) used in Chapter
\ref{qd_optics} have vanishing electrical field expectation
values. Nevertheless, the Bi-photon HOLOCAM can measure their wave
functions.


%

\end{document}